# Design and Implementation of a New Apparatus for Astrochemistry: Kinetic Measurements of the CH + OCS Reaction and Frequency Comb Spectroscopy in a Cold Uniform Supersonic Flow

Daniel I. Lucas,[a] Théo Guillaume,[a,†] Dwayne E. Heard,[b] and Julia H. Lehman[a],*

[a]School of Chemistry, University of Birmingham, Edgbaston, United Kingdom, B15 2TT

[b]School of Chemistry, University of Leeds, Leeds, United Kingdom, LS2 9JT

*Corresponding Author: Julia H. Lehman, j.lehman@bham.ac.uk

[†]Current address: LOMA, Université de Bordeaux, CNRS, UMR 5798, Talence Cedex, 33405, 351 cours de la Libération, FR

## Abstract

We present the development of a new astrochemical research tool HILTRAC, the Highly Instrumented Low Temperature ReAction Chamber. The instrument is based on a pulsed form of the CRESU (Cinétique de Réaction en Écoulement Supersonique Uniforme, meaning reaction kinetics in a uniform supersonic flow) apparatus, with the aim of collecting kinetics and spectroscopic information on gas phase chemical reactions important in interstellar space or planetary atmospheres. We discuss the apparatus design and its flexibility, the implementation of pulsed laser photolysis followed by laser induced fluorescence (PLP-LIF), and the first implementation of direct infrared frequency comb spectroscopy (DFCS) coupled to the uniform supersonic flow. Achievable flow temperatures range from 32(3) – 111(9) K, characterising a total of five Laval nozzles for use with $N_2$ and Ar buffer gases by pressure impact measurements. These results were further validated using LIF and DFCS measurements of the CH radical and OCS, respectively. Spectroscopic constants and linelists for OCS are reported for the $10^01$ band near 2890 – 2940 cm$^{-1}$ for both OC$^{32}$S and OC$^{34}$S, measured using DFCS. Additional peaks in the spectrum are tentatively assigned to the OCS-Ar complex. The first reaction rate coefficients for the CH + OCS reaction measured between 32(3) K and 58(5) K are reported. The reaction rate coefficient at 32(3) K was measured to be 3.9(4) × 10$^{-10}$ cm$^3$ molecule$^{-1}$ s$^{-1}$ and the reaction was found to exhibit no observable temperature dependence over this low temperature range.

# I. Introduction

For many decades, molecules, even those complex in nature,[1] have been known to exist in interstellar gas clouds despite the harsh conditions of the interstellar medium (ISM). Around 300 molecules have now been identified since the methylidyne radical, CH, was first observed in space in 1937.[2, 3] Among the ever-increasing number of molecules observed, there are ~28 sulfur containing molecules,[4] with recent detection of *trans*-HC(O)SH and HNSO toward the Galactic Center quiescent cloud G+0.693–0.027.[5, 6] Identification of such a diverse range of molecules in the ISM has been made possible due to the development of advanced astronomical instruments, such as the Atacama Large Millimeter/submillimeter Array (ALMA) in the Chilean desert and the James Webb Space Telescope (JWST).[7, 8] However, the formation and destruction pathways of such molecules, particularly for reactions of organosulfur species, at temperatures and gas densities ranging from 10 – 100 K and $10^2$ – $10^8$ molecule $cm^{-3}$, respectively, remain largely unknown. Large chemical reaction networks have been established to accurately account for the observed abundances of interstellar molecules, yet current astrochemical models cannot always accurately reproduce the observed molecular abundances.[9]

These discrepancies between models and observation are often caused by a lack of experimental data pertaining to the chemistry of the ISM, particularly parameters such as reaction rate coefficients and product branching fractions. Regarding S-bearing species, chemical networks are relatively incomplete when considering the formation and loss of such molecules. Cernicharo et al. evaluated the current knowledge for six S-bearing species using the UMIST RATE12 database with additional improvements based on the work of Vidal et al.[10-12] However, even this model underpredicted the abundances of some S-bearing species relative to the abundance $H_2$ by 1-2 orders of magnitude compared to observations, primarily due to the uncertain reaction rate coefficients in elementary reactions of atomic radicals with various S-bearing species. Many models of S-bearing species rely on the assumption that the chemistry of sulfur-containing and oxygen-containing species is similar due to the valence isoelectronic nature of sulfur and oxygen. This is evident in the recent release of the UMIST RATE22 database for astrochemistry where the photodissociation rate of HCS included in the database is the same as for HCO, and data pertaining to HSC is omitted from the database due to its gas-phase synthesis being unknown.[9] There is clearly an increasing need for experimentalists to develop state-of-the-art laboratory tools to drive the understanding and interpretation of observations through spectroscopic and kinetic measurements, as well as suggesting chemistry that could be missing from current models (networks) or new molecules to search for via observations.

When developing new experimental methods to study gas-phase chemical reactions important for the ISM, the first consideration is that the apparatus must be capable of generating a suitable medium at low temperatures (10 – 200 K). Methods such as cryogenic cooling of a flow cell reactor and supersonic adiabatic expansion (free-jet expansion) can generate cold gas flows satisfying the first criteria, but have major drawbacks for kinetic studies. Cryogenic cooling or heating of a flow cell reactor are capable of measuring temperature and pressure dependent reaction rate coefficients down to 200 K,[13-15] however, at lower temperatures, many molecules of astronomical interest can condense at the cell walls when the partial pressure of the reagent species is greater than their saturated vapour pressure. This impacts the ability to accurately measure reactant concentrations and thus limits the types of radical precursors and co-reagents that can be used in kinetics experiments. In contrast, the free-jet expansion of a high-pressure gas through a small orifice can generate temperatures on the order of 10 K or less and is widely used in spectroscopic studies.[16] While the expanding gas is cooled in the free-jet expansion, the gas is not uniform in temperature or density, making it very challenging to determine a reaction rate coefficient with this method. These limitations were addressed with the development of the CRESU technique (French acronym for Cinétique de Réaction en Écoulement Supersonique Uniforme, meaning reaction kinetics in a uniform supersonic flow) in the 1980s.[17, 18] The technique and its development has been extensively reviewed in articles and a recent book, and so only a brief description of the method follows.[19-21]

In the CRESU method, the supersonic, adiabatic expansion of a gas from a high-pressure reservoir into a low-pressure vacuum chamber is guided through a Laval nozzle that has a smooth, axisymmetric convergent-divergent internal profile. The result is a supersonic flow that is uniform in velocity, temperature, and gas density that persists for several tens of centimetres from the nozzle exit, which is a major advantage over other low temperature experimental techniques. This is essentially a wall-less reactor and ideally suited to the study of cold temperature chemical kinetics.[20-22] The original CRESU design requires a continuous flow of gas into the reservoir and through the expansion, which necessitates large gas flow rates (~30-90 SLM) and very high pumping capacities (~30,000 $m^3$ $h^{-1}$) in order to reach the lowest temperatures, down to 8 K.[17, 23-28] This prompted the development of a pulsed CRESU system,[29] which is currently or has been utilised by several groups around the world. [30-33]

A typical pulsed CRESU utilizes pulsed solenoid or piezostack valves to periodically fill a smaller reservoir (~1 – 20 $cm^3$) with gas, which enters through a small orifice (on the order of a few millimeters) and is emptied by expansion through the Laval nozzle. During the pulse, the pressure reaches a steady-state value inside the reservoir. An advantage of the pulsed CRESU technique is that the reduced gas flow rates (< 20 SLM), and hence gas consumption, which reduces the required pumping capacity to < 2,000 $m^3$ $h^{-1}$. A more recent hybrid of the pulsed and continuous CRESU apparatuses has been developed that essentially pulses the gas through the nozzle into the chamber by using an aerodynamic chopper wheel.[34-37] The main motivation of this development was to keep the smaller gas consumption and pumping capacity compared to the continuous CRESU, but to also reduce the turbulence in the expansion compared to the valve-driven pulsed CRESU by keeping the reservoir pressure and flow rate more stable. However, these systems are difficult to build and maintain compared to the use of pulsed solenoid or piezostack valves.

With the CRESU method able to create a suitable medium at low temperatures in which to study a range of chemical reactions,[19, 20, 22] a further consideration is that the apparatus must be coupled to a detection method, or even several detection methods, to measure reaction rate coefficients and product branching fractions for the reaction of interest. Pulsed laser photolysis followed by laser induced fluorescence (PLP-LIF) is one method that is ubiquitous in the field of gas-phase chemical kinetics.[38-44] The PLP-LIF method allows one to initiate a reaction, then probe a specific rovibronic transition of the molecule of interest (typically a radical reactant) and collect its fluorescence as a function of reaction time to obtain the relative concentration of the species of interest.[19] While LIF is highly selective for the molecule of interest, quantitative product branching fractions, although possible, are difficult to obtain from this method.[45-47] LIF is also fairly limited in its applications, with LIF schemes most often available only for small radicals. Other chemical kinetics detection methods coupled to a USF include chemiluminescence,[48] mass spectrometry,[30] and cavity ring-down spectroscopy.[49]

More recently, a chirped-pulse Fourier transform microwave spectrometer has been coupled to the USF.[33, 50-52] Microwave spectroscopy employed in this manner allows for a wide frequency range to be detected, collecting detailed spectroscopic and structural information about the molecules present in the USF. More than one species can potentially be monitored simultaneously, such as observing reaction products. This method of detection has the potential to allow for accurate product branching ratios from the relative spectral line intensities. The spectral bandwidth and frequency range could introduce limitations in these measurements, however, particularly if vibrationally excited species are produced in the reaction. The high collision rate within the USF also dampens and reduces the amplitude of the free-induction decay signal for these experiments, reducing the achievable signal-to-noise. Secondary expansion into a lower pressure environment has helped overcome the signal-to-noise limitation and is still being further developed.[53]

An alternative detection method coupled to the USF, and potentially quite promising, is that of direct frequency comb spectroscopy (DFCS).[54] It is still a broadband spectroscopic method and able to potentially monitor more than one species simultaneously, but now simply uses direct absorption spectroscopy via the Beer-Lambert Law, without having the same issues as chirped-pulse microwave spectroscopy with the high gas density. DFCS has been used successfully in a variety of gas-phase reaction kinetics studies, primarily in cavity-enhanced configurations surrounding flow cells or high temperature environments.[55-57] Its use in a USF has not yet been published to the best of our knowledge, although it is in development elsewhere in addition to the work presented here.[58] Translating this technology to USF studies has its challenges, such as mitigating against vibration induced by the high throughput vacuum pumps used in these studies. Infrared spectroscopy using single frequency lasers[49, 59] or broadband light sources[60] have already been demonstrated in a USF, using multipass, cavity ringdown, or cavity enhanced designs (see Chapter 9 of Ref 19 and references therein). Multipass configurations are less susceptible to vibrations, but at the expense of the reducing the pathlength through the USF, limiting the sensitivity of the measurement.

In this paper, we present the development of a new instrument for astrochemical research, the Highly Instrumented Low Temperature ReAction Chamber (HILTRAC). The major goal of HILTRAC is to develop a versatile and sensitive instrument with multiple detection techniques capable of measuring kinetic and spectroscopic parameters, particularly at low temperatures. In this article, we present the design and experimental setup of the newly commissioned HILTRAC apparatus (Section II) and we report the very first results from the newly commissioned astrochemical research tool (Section III). We discuss in detail the chamber design (Section II.A), the methods used to perform kinetic studies on the CH + OCS reaction (Section II.B), coupling of LIF to USF (Section II.C), and the first implementation of DFCS coupled to the USF (Section II.D). We then present the characterisation of the uniform supersonic flow (Section III.A), the infrared spectroscopy measurements resulting from coupling DFCS with the USF (Section III.B), and the first measurements of the CH + OCS reaction rate coefficients below 297 K (Section III.C).

## II. Experimental Methods

The HILTRAC apparatus is detailed in this section, with specifics given for the apparatus itself, specifics for the optical components for laser-induced fluorescence and frequency comb spectroscopy, and the experimental methodology for the kinetics experiment of CH + OCS. All manufacturer product codes for commercial instrumentation can be found in Table S1 of the supplementary material.

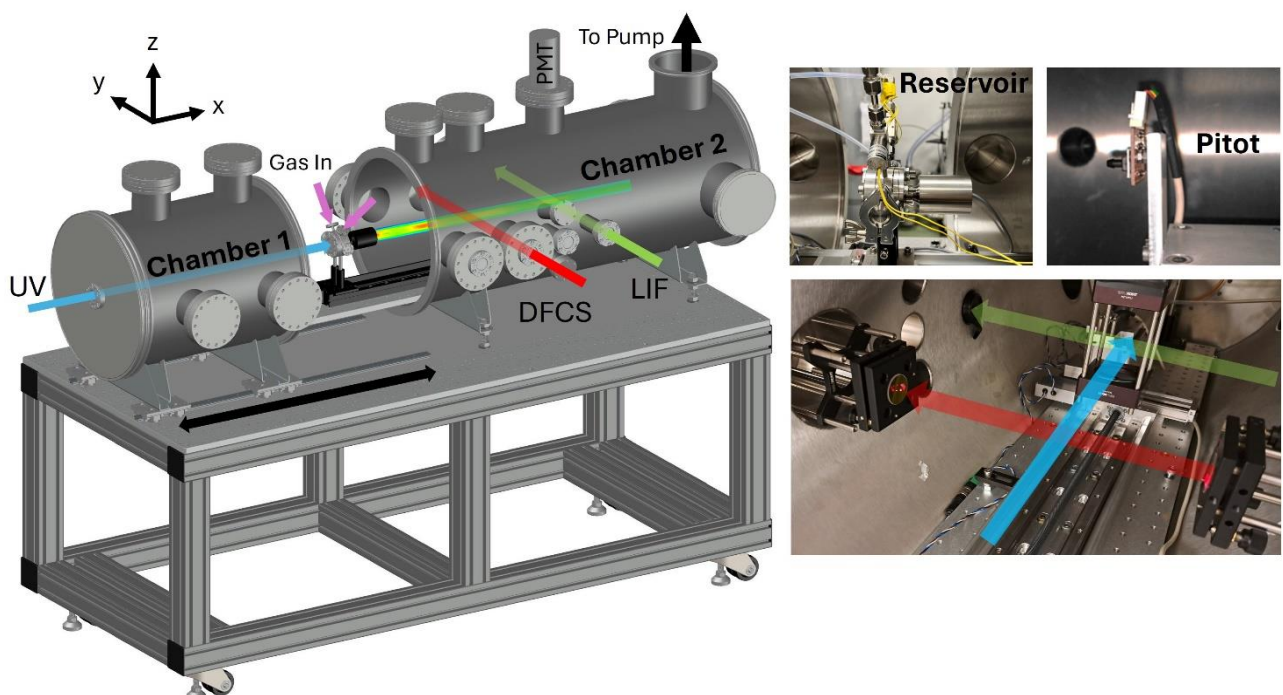

**Figure 1.** HILTRAC apparatus schematic and inset photos. The apparatus schematic highlights the moveable Chamber 1, the path of the UV photolysis laser (light blue), the reservoir with gas inlets (purple arrows), the USF produced from the nozzle, and the intersection of the flow with the DFCS (red) and LIF (green) optical axes. The top inset photos show the reservoir with Laval nozzle (left) and the Pitot tube mounted on the YZ translation stages (right). The bottom inset photo shows an interior picture of the chamber, with the X translation stage for the reservoir visible on the internal breadboard, the two optical axes for LIF and DFCS overlapping with the UV photolysis laser, and the cage-mounted LIF collection optics. A red diode laser shows the approximate path of the infrared frequency comb laser beam as it reflects off the gold mirrors for the Herriott multipass configuration. Further detail is in the text.

### A. Vacuum Chamber Apparatus

The HILTRAC experimental apparatus (Figure 1) consists of two custom vacuum chambers (Kurt J. Lesker) mounted on an aluminium extrusion frame topped with a breadboard. The first vacuum chamber is moveable relative to the second chamber, making it straightforward to access the inside of the apparatus to, for example, change Laval nozzle. The two chambers are 40 cm in diameter, with the first chamber being 0.5 m long and the second chamber being 1 m long. The first chamber has ports with vacuum flanges (Lewvac) that allow for: numerous electronics feedthroughs, optical access by the excimer laser (see below), gas feedthroughs to the reservoir, optional slip gas leak valves, and 2 Torr and 1000 Torr Baratron capacitance manometer gauges (MKS) for measuring chamber pressure and calibrating other pressure transducers. The stationary second chamber has ports and custom flanges to allow optical access to the supersonic flow. These optical arrangements are described further in the next section, but briefly consist of optical access and detection for laser-induced fluorescence studies and infrared laser access for frequency comb spectroscopy. The last port on the top of the second chamber is used for the high throughput vacuum pump (Edwards GXS 250/2600, max. 1900 m$^3$ h$^{-1}$). Since the apparatus is purposefully designed to be highly flexible, additional blank flanges are currently on the apparatus as well, allowing for future modifications to house alternative instrument access to the supersonic flow.

A 1.4 m × 0.2 m breadboard is welded to the inside of the second chamber, spanning the full length of the two chambers but not moving when the first vacuum chamber slides backwards on its track. Mounted on the breadboard are vacuum-compatible translation stages (Zaber) which translate the reservoir along the length of the chamber (X-axis) and translate a Pitot tube in the YZ-plane. Positioning of the reservoir and Pitot tube is achieved via an electronic interface, with the recirculating ball bearings with precision lead screw drive system of the stages achieving micrometer resolution and absolute distance resolution. The reservoir and nozzle can move up to 60 cm in the X-dimension. The pitot tube is mounted on the stacked pair of vertical and horizontal translation stages using an L-shaped bracket, as seen in Figure 1. The manufacturer accuracy and resolution specifications are given in Table S2. All three stages are controlled via a LabVIEW programme, and the "zero" position of the pitot tube relative to the central axis of the reservoir and nozzle is determined via laser alignment, and can be confirmed by collecting 2D contour plots of the uniform flow.

The reservoir (Figure 1) is based on a custom DN40CF flange (Lewvac), with a demountable viewport on one side to allow excimer (photolysis) laser access and a blank flange on the front to mount the Laval nozzle. The blank flange allows for Laval nozzles to be attached to the reservoir with an o-ring seal, either using a clamping mechanism around the exterior of the nozzle, or by screwing the nozzle into the threaded tapped hole on the flange. Two gas feedthroughs (6 mm outer diameter tube) and two monitoring ports (KF16, for pressure gauge and optional thermocouple) are arranged at 60° spacing around the exterior of the central DN40CF flange. The total internal volume, including the feedthrough ports, is 35.1 cm$^3$. This is larger than the previous pulsed valve reservoirs of the Heard and Suits groups,[32, 33] and uses two pulsed solenoid valves (Parker) with large orifice sizes (2.95 mm) to achieve a steady state flow into and out of the reservoir resulting in a constant pressure in the reservoir. An example of the reservoir pressure as a function of time before, during, and after the pulsed valves fire is given in the supplementary material (Figure S5). The reservoir pressure is measured by a fast pressure transducer (Omega), which is regularly calibrated against the capacitance manometers mentioned above. The pulse width and relative timing of the two pulsed valves are controlled by a custom-built pulsed valve driver. Gas is handled by a mixing manifold with Tylan and MKS flow controllers and a custom flow controller electronic interface to set and read out the flow rates with computer control. For a typical kinetic experiment, separate flow controllers are used to control the amounts of an excess reagent, precursor molecule (often a liquid whose vapour is entrained in a buffer gas via a bubbler), and a further buffer gas flow to balance the total density. The gasses enter a pre-expansion reservoir (known as the "ballast") to allow for sufficient mixing and maintaining a constant pressure behind the pulsed valves.

The Pitot tube (Honeywell) measures the impact pressure of the supersonic flow achieved with the Laval nozzle, allowing for the performance characterization of each Laval nozzle which is discussed further in the Results section. The Pitot tube is regularly calibrated using the capacitance manometers mentioned above. The 3D translation of the Pitot tube relative to the Laval nozzle allows for 1D, 2D, or 3D maps of the achievable flow conditions, as discussed further in the Results section. To control the reservoir and chamber pressures to achieve optimal flow conditions, alterations can be made in the relative timing of the two pulsed valves, gas flow rates, the addition of a slip gas flow into the chamber, and vacuum pumping speed. An optimal flow is one in which there are minimal variations in the temperature and number density as a function of the distance from the nozzle exit, and where the longest distance maintaining uniformity of the flow is observed. Measurements of the length of uniform flow help to set the maximum distance between the nozzle exit and the laser interaction region during kinetics studies, and will vary depending on the nozzle. Information on operating conditions for various nozzles are given in the Results section.

## B. Laser-Induced Fluorescence

Three ports on the second chamber are dedicated to LIF studies, with two ports for laser access and a third, perpendicular port containing flange-mounted LIF collection optics and photomultiplier tube (PMT) detector. The "probe" laser used for CH LIF studies is the frequency doubled output from a dye laser (Sirah, LDS722 dye) pumped by the 532 nm output of an Nd:YAG (Continuum 8000) laser. A laser spot size of 2 mm and pulse energy of 2 mJ is typically used. The UV probe laser is oriented perpendicular to the flow direction and is spatially overlapped with the uniform flow and excimer laser. It is used to excite the CH B $^2\Sigma^-$ ← X $^2\Pi$ (1,0) Q$_2$(1) transition at ~363.432 nm, and fluorescence is detected around 400 nm on the B $^2\Sigma^-$ → X $^2\Pi$ (1, 1) transition. The fluorescence detection axis is perpendicular to the plane made by the excimer photolysis laser and UV probe laser. A concave rear reflector and biconvex lens surround the interaction region and a further lens focusses the light past a bandpass filter (Semrock, 405 nm, 10 nm bandwidth) and into a PMT (ET Enterprises). The LIF collection optics are cage mounted (Thorlabs) to the PMT flange to allow for the distances between optics and the interaction region to be optimized for the largest fluorescence signal, as well as ensuring the mechanical alignment with the flange center and chamber axis. The LIF signal is processed with an oscilloscope (LeCroy WaveRunner 6100) and transferred to a laboratory computer for further analysis. A single exponential decay function is fit to the LIF decay data and integrated to obtain a value proportional to the relative amount of CH in the flow. The CH fluorescence lifetime is unchanged during the experiments described here, measured to be approximately 450 ns. The primary collision partner in the expansion is argon, which has a very small electronic quenching cross section for the B $^2\Sigma^-$ state of CH,[61, 62] and so it is unsurprising that the measured lifetime is comparable to the natural lifetime of CH B $^2\Sigma^-$ (v = 1).

## C. Direct Frequency Comb Spectroscopy

Two sets of optical ports on the HILTRAC apparatus are available for the frequency comb laser to intersect the uniform flow perpendicular to the flow direction, one allowing for a Herriott multi-pass configuration and one in a cavity-enhanced configuration. The cavity-enhanced configuration is part of ongoing work and will be discussed in a future publication. The Herriott multi-pass configuration is presented in this work instead. The frequency comb laser (Menlo Systems, Mid-IR Comb 3-5) operates over approximately 3000 – 5000 nm (2000 – 3300 cm$^{-1}$) with a tuneable 200 nm bandwidth. For the experiments discussed here, the comb was primarily used in the 3200 – 3400 nm range. The femtosecond frequency comb laser has nominally a 250 MHz pulse repetition rate and is frequency referenced to a rubidium oscillator (SRS FS725). The repetition rate is tuneable, allowing for the frequencies of the comb to be scanned, resulting in an interleaved dataset with a finer grid of the frequency datapoint spacing than otherwise possible with a single repetition rate. This is particularly useful at low temperatures or in other instances where the lineshape function of the molecular transition, attributed mainly to Doppler broadening in these experiments, is narrower that the nominal spacing between comb teeth (250 MHz or 0.008 cm$^{-1}$).

The laser is fiber coupled to and from the HILTRAC apparatus. Herriott mirrors (Thorlabs) are mounted interior to the chamber using flange-mounted Thorlabs cage components, and wedged calcium fluoride windows (Crystran) allow for light to enter and exit the chamber. The Herriott multipass configuration with off-axis entrance and exit holes allows for 23 passes through the uniform flow. A reverse-propagating red diode laser assists in the alignment of the mirrors and the number of passes of the visible alignment laser is assumed to be the same as the number of passes of the infrared light. This assumption is further validated by the derived pathlength based on molecular absorption and Pitot measurements of the uniform flow core size, as discussed in the Results section.

After traversing the uniform flow, the laser is fiber coupled to the spatially dispersive spectrometer. The comb spectrometer in our research group has previously been discussed,[63] and so only a brief description follows and highlights the changes made. Following fiber coupling, the collimated laser is focussed using a cylindrical lens into a virtually-imaged phased array (VIPA, LightMachinery). The new VIPA has a prism-based coupling design, which allows for higher coupling efficiency at shallower tilt angles of the optic with respect to the incoming laser beam. The VIPA has an FSR of approximately 9 GHz, as opposed to the 13 GHz VIPA used by Roberts et al.[63] The vertically dispersed light is then cross-dispersed using a diffraction grating, and then imaged on a position sensitive InSb array camera (Infratec, 640 × 514 pixels). Given the fluctuations in laboratory temperature and the sensitivity of the VIPA to any changes in temperature, a thermoelectric cooler with PID temperature controller holds the VIPA at 19.0°C.

For the pulsed experiments performed here, the frequency comb data collection is operated in an active background subtraction mode where the camera is synchronized to the gas pulses in the same manner as the LIF probe laser above, but the camera is triggered at 20 Hz while the experiment is triggered at 10 Hz. This means that every other camera image has either the gas sample present (signal, I) or not present (background, $I_0$) in the interaction region between the gas flow and the laser beam. A similar scheme has been utilized and characterized for a pulsed molecular beam in our group, the details of which can be found elsewhere.[64] The camera integration time (frame exposure time) is 10 μs, and a total of 4000 images (2000 signal, 2000 background) are recorded for each comb repetition rate. The resulting signal and background images are transformed into infrared absorption spectra using the Beer-Lambert Law, where A = log($I_0$/I). The repetition rate of the comb is stepped by 50 Hz, from 249.999700 MHz to 250.000350 MHz. The absorption spectra recorded at each repetition rate are interleaved, resulting in a datapoint spacing of approximately 0.0006 cm$^{-1}$ in the final infrared spectrum reported here. A second experiment is performed at a single repetition rate (250.000000 MHz) with only argon in the uniform flow, allowing for the achievable noise level of the new apparatus to be tested at both 10 μs and 2 μs of camera integration time, which is discussed in the supplementary material.

### D. Kinetics Experiment Methodology

An overview of the experimental methodology for measuring temperature dependent reaction rate coefficients for the CH + OCS reaction is as follows: a gas mixture containing $CHBr_3$ (Sigma Aldrich 99%), OCS (BOC 20.00% calibrated mix in argon), and an argon buffer gas are pulsed into the reservoir and through a Laval nozzle, generating a uniform supersonic flow with typical densities of $CHBr_3$ ~$4.1 \times 10^{12}$ molecule $cm^{-3}$, OCS $0.3 - 12.9 \times 10^{13}$ molecule $cm^{-3}$, and the buffer gas balancing to a total density of $4.3 \times 10^{16}$ molecule $cm^{-3}$ at 32(3) K. A KrF excimer laser (Coherent COMPex 102) is directed coaxially with the uniform flow, photolysing $CHBr_3$ at 248 nm (4 mm spot size, 70 mJ per pulse after the Laval nozzle) to generate CH radicals.[42] The OCS concentration is held in excess, with the CH + OCS reaction progressing under pseudo-first order conditions. Approximately 10 – 20 cm downstream from the exit of the nozzle, the probe laser (UV LIF excitation laser) intersects the uniform flow and the resulting laser-induced fluorescence signal is used to monitor the relative amount of CH in the flow. By changing the time delay between the excimer (pump) and UV LIF probe lasers, a kinetic time trace of $[CH]_{relative}$ versus time can be made for a specific OCS excess concentration. The OCS concentration is then changed, with the argon buffer gas concentration being reduced or increased to compensate and maintain the uniform flow conditions, and the experiment is repeated. As discussed in detail in Section III.D, a line is fit to a plot of the pseudo first order reaction rate coefficient versus OCS concentration, where the slope is the second order reaction rate coefficient. This second order reaction rate coefficient is for the specific temperature generated by the Laval nozzle used for the uniform supersonic flow, as derived from Pitot tube impact pressure measurements. The Laval nozzle is then changed to a different nozzle profile yielding a different temperature, and the experiment is repeated to derive the reaction rate coefficient at this new temperature.

The experimental timings of the pulsed valves, excimer laser, and UV LIF probe laser are controlled by a digital delay generator (DDG, Berkeley Nucleonics Corporation, Model 565), which triggers the experiment at 10 Hz. The PMT gate is triggered using the DDG, with the gate width being 4 μs and relative timing set such that any excimer laser scattered light observed by the PMT is minimized. The DDG, reservoir translation stage position, pressure readings, gas flow controllers, and LIF signal collection and processing are all controlled and automated in LabVIEW. The frequency comb spectrometer is also triggered by the DDG, with the data collection of the comb spectrometer separated from the main kinetics programme and detailed above. Two photodiodes can be used to monitor the power of the photolysis and probe laser beams to account for fluctuations in the laser power.

## III. Results and Discussion

### A. Characterization of the Uniform Flow

A total of five Laval nozzles were designed for this work, summarised in Table 1, with four of the nozzles being replicas of those used by the Heard research group.[32] The naming convention for the nozzles in our group is based on the order they were printed or machined. Nozzle 9 is a machined nozzle in stainless steel, a replica of the M4 nozzle in use by the Heard group.[32] Printed nozzles 2-4 are 3D printed Laval nozzles in GreyPro resin (FormLabs) and are replicas of the M3.3, M2.5, and M2.25 nozzles in use by the Heard group. The fifth nozzle is also a 3D printed nozzle in black resin that was newly designed for this work. Before characterising the USF, both the reservoir pressure transducer and the Pitot tube were calibrated against the 2 Torr and 1000 Torr capacitance manometers. Here, vacuum is taken as the base pressure and the sensors are calibrated up to the full range of the reservoir pressure transducer of 350 mbar. Representative calibration curves can be found in Figure S1 of Section SII of the supplementary material.

Following these calibrations, the USF properties generated by each of the nozzles were characterised by pressure impact measurements using a Pitot tube. The Pitot tube (mounted as discussed in Section II.A) is placed at the center of the flow axis to measure the impact pressure ($P_i$) of the USF as a function of displacement in the three dimensions, defined as axial (X), radial (Y), and vertical (Z) as shown in Figure 1. The nozzle and Pitot tube are positioned along the vertical and radial center of the chamber using laser alignment. The nozzle position within the chamber was set such that the exit plane of the Laval nozzle is at the face of the Pitot tube at X = 0 cm. The Pitot tube was set to be in the center of the USF and at the height of the LIF axis at Y = 0 cm and Z = 0 cm, respectively. An uncertainty in the "zero" position of the Pitot tube would conservatively be on the order of 0.5 mm. Finally, an uncertainty analysis was performed to determine the error that contributes to the overall uncertainty in the flow properties.

To determine the flow properties from pressure impact measurements, the experimental Mach number must first be determined by iteratively solving the Rayleigh Pitot equation (Eq 1),[21] where the Mach number ($M$) is optimised to match the observed ratio of the impact pressure to reservoir pressure ($P_i/P_0$) at each displacement value. The heat capacity ratio ($\gamma = C_p/C_v$) is known for common buffer gasses ($\gamma$ = 1.6667 for Ar or $\gamma$ = 1.4 for $N_2$).

$$\frac{P_i}{P_0} = \left(\frac{(\gamma+1)M^2}{(\gamma-1)M^2+2}\right)^{\frac{\gamma}{\gamma-1}} \left(\frac{\gamma+1}{2\gamma M^2-\gamma+1}\right)^{\frac{1}{\gamma-1}} \qquad (1)$$

Once the Mach number has been calculated, the temperature ($T_1$), pressure ($P_1$) and gas density ($n_1$) can all be calculated using the thermodynamic relationships given in Equations 2-4.

$$\frac{T_0}{T_1} = 1 + \frac{\gamma-1}{2}M^2 \qquad (2)$$

$$\frac{P_0}{P_1} = \left(\frac{T_0}{T_1}\right)^{\frac{\gamma}{\gamma-1}} \qquad (3)$$

$$\frac{n_0}{n_1} = \left(\frac{T_0}{T_1}\right)^{\frac{1}{\gamma-1}} \qquad (4)$$

**Table 1.** Parameters of the USF for each Laval nozzle used in this work in Ar and $N_2$, characterised by pressure impact measurements. The experimental values are the reservoir pressure ($P_{res}$) and chamber pressure ($P_{ch}$). The calculated USF properties include the maximum flow length (L), corresponding maximum kinetic time, Mach number, flow pressure (P), flow temperature (T), and flow number density (D).

| Experimental Parameters | | | | Calculated USF Properties | | | | | |
|---|---|---|---|---|---|---|---|---|---|
| Buffer Gas | Nozzle | $P_{res}$ (mbar) | $P_{ch}$ (mbar) | L (cm) | Kinetic Time[a] (μs) | Mach | P (mbar) | T (K) | D ($10^{16}$ molecule cm$^{-3}$) |
| Ar | 9 | 48.3 | 0.23 | 26.0 | 498 | 4.9(2) | 0.2(1) | 32(3) | 4(1) |
| | P2 | 95.1 | 1.27 | 23.0 | 448 | 4.2(4) | 0.9(4) | 44(7) | 14(3) |
| | P3 | 56.0 | 1.16 | 16.0 | 320 | 3.7(2) | 0.8(2) | 54(5) | 11(2) |
| | P4 | 54.0 | 1.40 | 14.0 | 282 | 3.5(2) | 0.9(2) | 58(5) | 12(1) |
| | P5 | 50.1 | 1.05 | 15.0 | 296 | 3.9(2) | 0.5(1) | 48(5) | 8(1) |
| $N_2$ | 9 | 55.8 | 0.38 | 33.0 | 483 | 4.1(1) | 0.3(1) | 68(2) | 4(1) |
| | P2 | 92.6 | 1.69 | 23.0 | 353 | 3.4(2) | 1.5(5) | 89(7) | 12(3) |
| | P3 | 53.3 | 1.28 | 18.0 | 286 | 3.1(1) | 1.3(4) | 102(6) | 9(1) |
| | P4 | 52.2 | 1.62 | 15.0 | 239 | 2.9(1) | 1.8(5) | 111(9) | 11(2) |

[a] An example calculation is given in Section III.D of the main text.

The 2D temperature and gas density profile for the USF generated by Nozzle 9 is shown in Figure 2 for an argon buffer gas. Here, the two solenoid valves were pulsed at 10 Hz, with 10 ms pulse widths and the relative timings adjusted to give a stable reservoir pulse for ~10 ms. The reservoir and chamber pressures were set to 48.3 mbar and 0.23 mbar, respectively. The reservoir was moved away from the Pitot tube in 0.2 cm increments along the axis of the flow to increase the distance between the Laval nozzle exit and the Pitot tube, while the Pitot tube was displaced horizontally across the USF from -1.8 – 1.8 cm in 0.2 cm intervals at the height of the center of the flow axis. Thus, resulting in a 2D image of the XY-plane of the USF. At each displacement, the reservoir pressure and impact pressure values across the central 2 ms of the pulses were averaged 20 times, which is discussed further in the supplementary material (Section SIV). The average temperature and gas density over the maximum distance that the flow maintained uniformity was measured to be 32(3) K and 4(1) × $10^{16}$ molecule cm$^{-3}$, respectively. The uncertainty represents the variation in the flow temperature or density profile of the maximum distance of uniformity with further discussion on error propagation given in Section SIV of the supplementary material. The maximum distance of uniformity was conservatively quantified by finding the distance where the temperature of the USF fell below one standard deviation of the mean value over the first 10 cm of the USF. Additional 2D cross-sections of the USF using Nozzle 9 can be found in the supplementary material. Also included in the supplementary materials are the characterisation of the $N_2$ flow for the same nozzles. All other nozzles have been validated in one dimension using both $N_2$ and Ar as buffer gasses and these profiles can be found in the supplementary material.

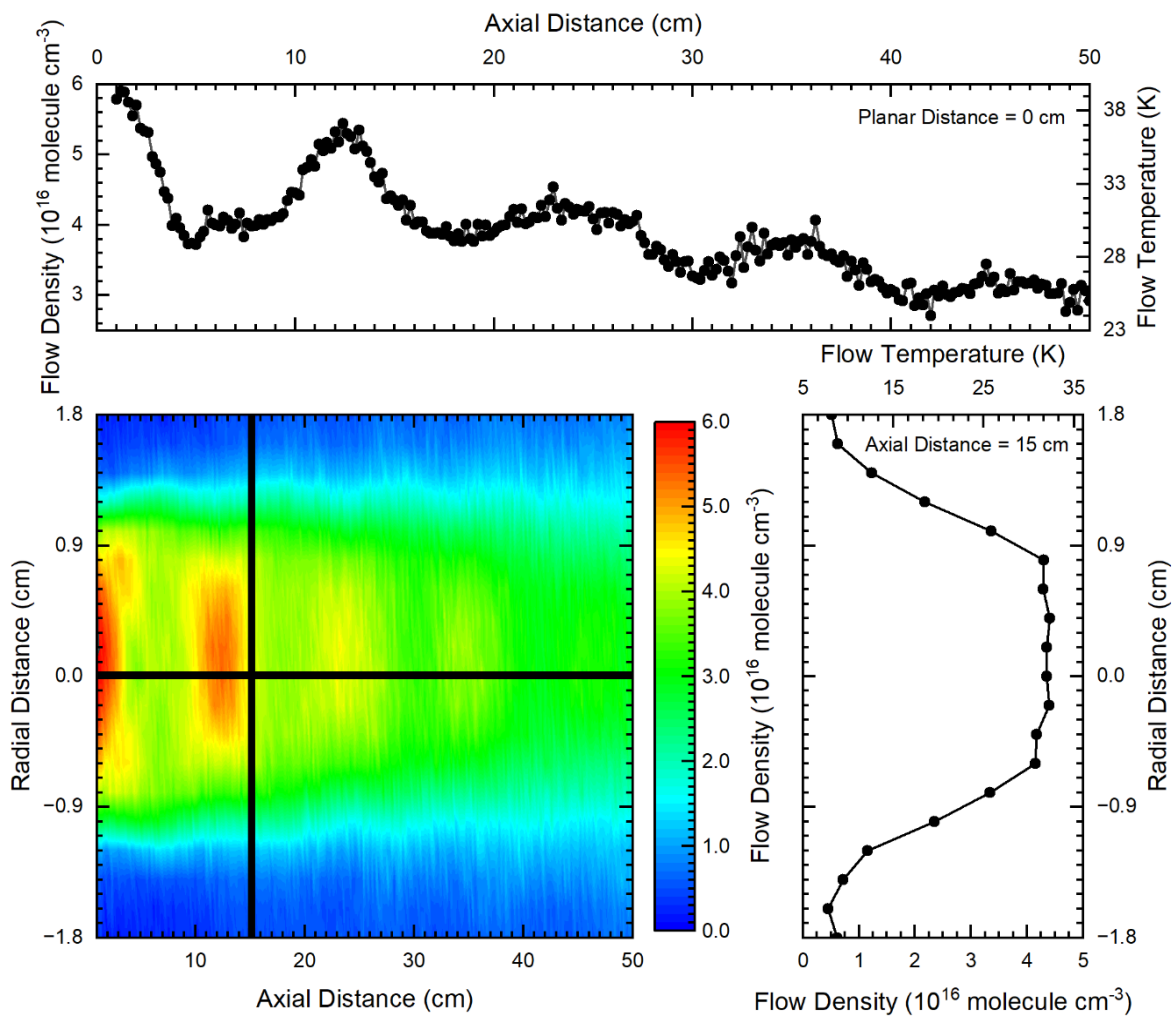

**Figure 2.** Contour map (bottom left) of the USF in an Ar buffer gas for Nozzle 9 from 1 – 50 cm in X and -1.8 – 1.8 cm in Y in 0.2 cm intervals in both cases. Also shown is a 1D profile in the X plane at Y = 0 cm (top) and a 1D profile in the Y plane at X = 15 cm (bottom right).

Figure 2 shows two features: oscillations in the flow temperature and gas density, and a decrease in the flow temperature and gas density along the axis of the flow. The uniformity of the supersonic flow was optimised by systematically changing the chamber pressure to reduce oscillations in the properties of the flow while maximising the length of uniformity. In general, the chamber pressure should match the flow pressure calculated in Equation 3 to achieve the longest and least oscillatory USF. Oscillations in the USF flow, which are indicative of shockwaves, are typical observations in all CRESU systems and are reproduced by computational fluid dynamics simulations.[19] Experimental variables such as reservoir pressure, chamber pressure, and nozzle shape all contribute to reduce the magnitude and frequency of the oscillations. In this apparatus, chamber pressure and reservoir pressure, which is influenced by valve pulse widths, valve delays, and flow rate, were finely controlled to achieve the most stable USF. Complementary computational fluid dynamics simulations are currently underway to fully characterize and optimize the flow conditions and nozzle design.[65] Eventually, the uniformity of the flow breaks down, invalidating the Rayleigh-Pitot equations. For kinetics purposes, it is necessary to know when the uniform temperature and density conditions are no longer maintained. The breakdown of the USF has been conservatively quantified as the distance at which the temperature falls below one standard deviation of the average temperature across the first 10 cm of the USF. The range of temperatures and gas densities present in the USF impact the kinetic measurements and are therefore considered when calculating the uncertainty in the reaction rate coefficients measured in this work.

## B. Rotational Temperature using Laser-Induced Fluorescence Spectroscopy

Whist the previous section used impact pressure measurements to derive the flow temperature, this section uses the rotational temperature measured using laser spectroscopy. The temperature measured in this way assumes that a Boltzmann distribution of rotational population is a direct reflection of the overall flow temperature. This is an oversimplification, since rotational and vibrational degrees of freedom will be collisionally cooled at different rates, especially by different collision partners. However, past work in the CRESU community has shown that the rotational temperature and the Pitot impact pressure measurements (which gives translational temperature) are generally consistent with one another.[32, 66-68] In this section, we discuss the measurement of rotational temperature using laser-induced fluorescence spectroscopy, while the next section shows results using infrared DFCS.

At a fixed distance downstream of the nozzle exit, the laser excitation wavelength was scanned and the integrated fluorescence signal recorded, averaging five fluorescence decay profiles per datapoint. Figure 3 shows an example laser excitation spectrum. The LIF regime (saturated, partially saturated, or linear) plays an important role in determining relative peak intensities and widths.[69] The spectrum shown here was taken under the linear regime. The UV laser power was monitored using a photodiode during all LIF spectra scans, changing by less than 10% during a scan. A PGOPHER simulation of the LIF spectrum was made with known CH rovibrational constants.[70, 71] Using contour fitting in PGOPHER, the LIF spectrum was fit to a rotational temperature. The resulting temperature of the coldest nozzle, Nozzle 9 (Ar buffer gas), measured 15 cm downstream of the nozzle exit was 40(2) K, with the reported error coming from the PGOPHER contour fit. The warmest nozzle (Print 4 in $N_2$ buffer gas) was also fit using PGOPHER, resulting in a rotational temperature of 128(5) K measured 15 cm downstream of the nozzle exit. The impact pressure measurements for these two nozzle and buffer gas combinations showed 32 K and 111 K at 15 cm downstream, which is coincidentally also the USF average temperatures of 32(3) K and 111(9) K. The measured temperatures are similar to the rotational temperature, although the LIF temperatures are generally higher than impact pressure measurements.

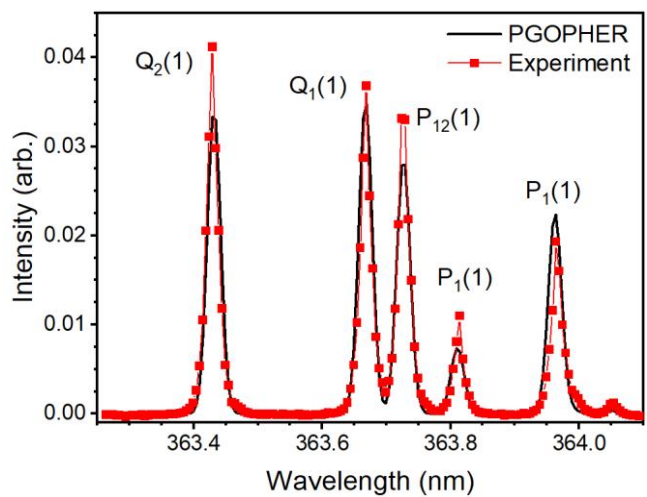

**Figure 3.** Experimental (red) LIF spectrum of CH B $^2\Sigma^-$ - X $^2\Pi$ (1, 0), with fluorescence detected around 400 nm on the (1, 1) transition. A PGOPHER simulation is overlaid (black) and used to fit the observed rotational temperature to a temperature of 40(2) K. Note that the $Q_2(1)$ transition is primarily used in the reported kinetics studies.

## C. Direct Frequency Comb Spectroscopy in a Uniform Supersonic Flow

This section will first discuss the frequency comb spectrometer performance when operating in an active background subtraction mode with a Herriott multipass configuration, and then discuss the interpretation of the observed infrared spectrum of OCS. A further discussion of the imaging conversion to spectra, noise analysis, and tabulated results from OCS spectral analysis is presented in the supplementary material.

To illustrate the capabilities of the frequency comb spectrometer, the OCS spectrum was measured using Nozzle 9 in an argon buffer gas, without any $CHBr_3$ added to the USF. The concentration of OCS in the flow was $5.49 \times 10^{13}$ molecule cm$^{-3}$. Although the frequency comb laser covers a wide (> 200 cm$^{-1}$) region of the infrared near 3000 cm$^{-1}$, the spectrometer used here only collects approximately 45 cm$^{-1}$ in a single image due to the field of view and size of the camera array. The angle of the diffraction grating can be changed and several images captured, as showcased previously,[72, 73] but this procedure was not necessary for the current study. The spectral data presented here cover from 2896.97 cm$^{-1}$ to 2940.13 cm$^{-1}$, which is the majority of the camera image. An example camera image is shown in Figure S10 of the supplementary material. The full spectrum is also shown in Figure S12 of the supplementary material compared to a HITRAN simulation.[74] Figure 4, panel A, shows a slightly truncated experimental spectrum (black) compared to the PGOPHER simulation (red, see details below),[70] with panel C showing the spectrum when using a single repetition rate (blue). It is immediately striking that the data collected at a single repetition rate are too sparse to accurately capture the full spectral profile. For example, the peak near 2918.9 cm$^{-1}$ is entirely missing from the single repetition rate spectrum, while the absorbance near 2919.3 cm$^{-1}$ is observed. It is necessary to take repeat measurements at different repetition rates and then interleave the data to obtain the final OCS absorption spectrum.

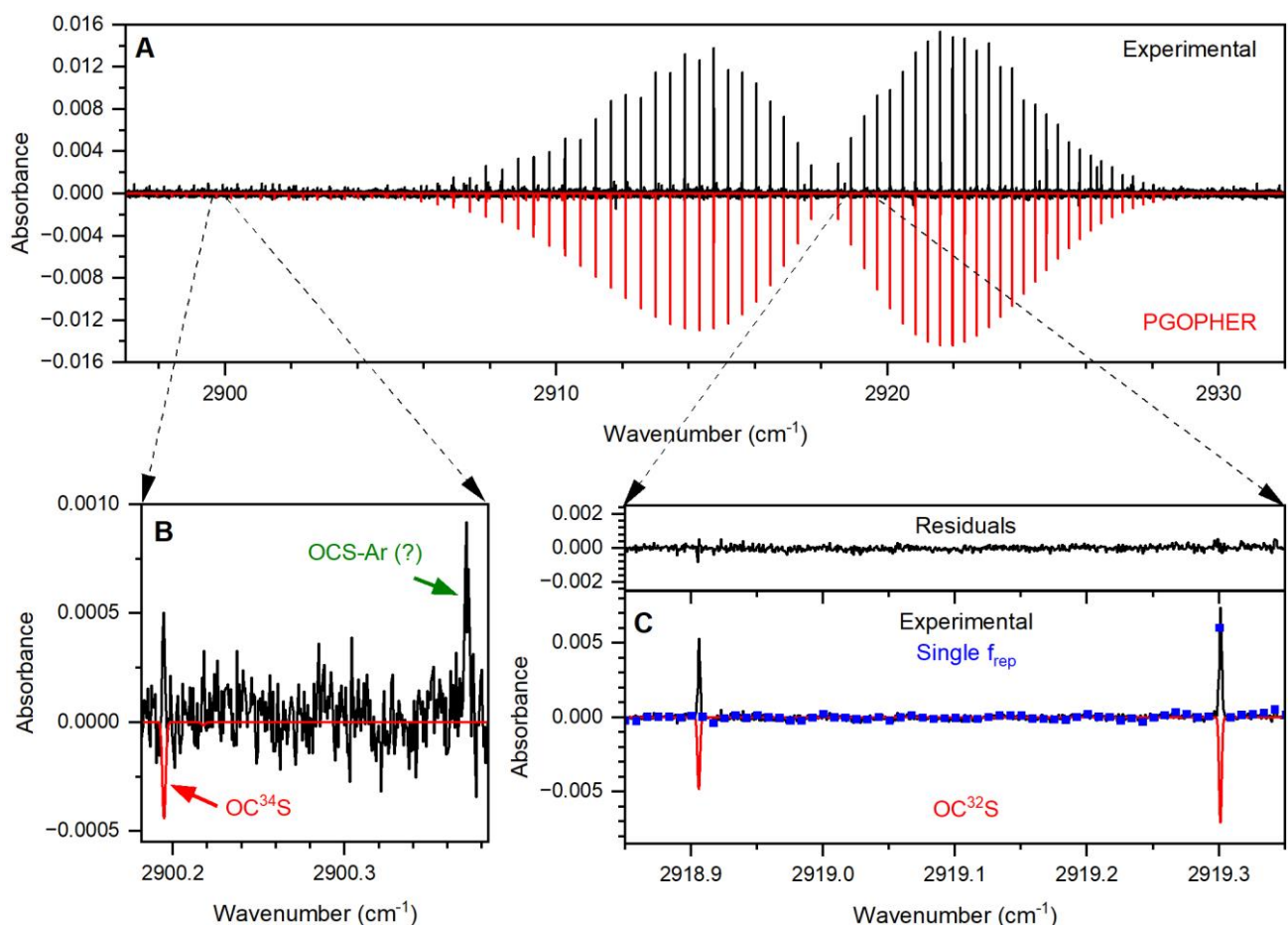

**Figure 4.** Infrared spectrum of OCS from 2897 – 2932 cm$^{-1}$ (black, panel A) compared against a PGOPHER simulation (red, panel A) using a temperature of 49.5 K and a Gaussian broadening FWHM of 0.00223 cm$^{-1}$. The inset in panel B shows a narrow wavenumber region near 2900 cm$^{-1}$, highlighting absorption features due to OC$^{34}$S and potentially OCS-Ar, discussed further in the text. While the experimental data in black in all three panels is the final interleaved spectrum, the spectrum collected at a single repetition rate (249.999800 MHz) is shown in panel C, with the blue symbols showing the discrete measured absorbances at each comb tooth. Also shown in panel C are the residuals between the experiment and the PGOPHER simulation. Note that all red traces (PGOPHER simulation) are inverted for clarity.

The standard deviation of the baseline of a single repetition rate spectrum (see Section SVII of the supplementary material for details), representative of the noise level, is $1.0 \times 10^{-4}$ when using 4000 images (2000 signal, 2000 background) as in the spectrum shown in Figure 4. The interleaved spectrum has a slightly higher noise level ($1.25 \times 10^{-4}$), and obviously takes longer to collect in the lab. For a 20 Hz data collection rate and 14 different frequency comb repetition rates with 4000 images, it took approximately 50 minutes to collect the data for the final spectrum consisting of 72,450 datapoints spaced by approximately 0.0006 cm$^{-1}$. While scanning the repetition rate was necessary to collect an accurate OCS infrared spectrum due to its narrow spectral lines, using this spectrometer for kinetic purposes only (not for detailed spectral fitting or interpretation) would not necessarily require multiple repetition rates to be used. If a single repetition rate were used with 4000 images using active background subtraction, then 5175 datapoints at 0.008 cm$^{-1}$ spacing and $1.0 \times 10^{-4}$ noise level could be collected in under 4 minutes.

The noise level can be further interpreted once the sample pathlength is determined. Using the known OCS line strength (HITRAN), the integrated absorption of 42 observed experimental peaks in the spectrum, and the concentration of OCS in the flow ($5.49 \times 10^{13}$ molecule cm$^{-3}$), the pathlength through the sample was found to be 31(2) cm. Given 23 passes determined by the overlapping visible laser beam, the pathlength corresponds to a 1.36(9) cm diameter USF. This is in excellent agreement with the core of the USF identified in Figure 2, where the flat-topped part of the density profile indicates a core of approximately 1.4 cm. Using the Beer-Lamber Law, the noise-equivalent value of σN (absorption cross section multiplied by the number density), which is the absorbance divided by the pathlength, is $3.2 \times 10^{-6}$ cm$^{-1}$. A discussion in terms of a 1-second noise-equivalent absorption is given in the SI. In the kinetics experiments carried out in a USF, typical concentrations of excess reactants are on the order of $10^{13}$ - $10^{14}$, while concentrations of non-excess reactants are potentially 2 orders of magnitude less. It would be highly unlikely to study non-excess reactants or products in the flow using a simple multipass design with only a moderate number of passes through the flow. For example, the largest peak observed in the experimental OCS spectrum corresponds to an integrated line intensity (S, in HITRAN) of $4.57 \times 10^{-20}$ cm molecule$^{-1}$, while some of the smallest peaks observed are on the order of $1 \times 10^{-21}$ cm molecule$^{-1}$. A reactant with a concentration of $10^{11}$ molecule cm$^{-3}$ would need a peak with an absorption cross section of at least $9 \times 10^{-17}$ cm$^2$ molecule$^{-1}$ for a signal-to-noise ratio of three, which is unlikely. At low temperatures, particularly strong absorbers in this region of the infrared, like methane, hydrogen cyanide, acetylene, and so on, would be able to be observed with this simple optical arrangement. A cavity-enhanced implementation is currently underway to help increase the pathlength through the flow. However, if kinetics are not the goal of the DFCS experiment and instead only spectroscopy of molecules under temperature-controlled conditions, the multipass design would be sufficient in most cases and offers benefits in its robust (relatively vibration-insensitive) design and a higher light intensity which is useful for the spatially dispersive spectrometer.

Apart from characterizing the spectrometer sensitivity, our aim is to measure and spectroscopically characterize high resolution infrared spectra collected with DFCS when coupled to a USF. In this region of the infrared, the OCS spectrum primarily shows the transition from the ground vibrational state to the combination band $10^01$, with contributions from the OC$^{32}$S and OC$^{34}$S isotopes (93.74% and 4.16% natural abundance, respectively). The OC$^{33}$S isotope is also expected to be present, but its natural abundance is approximately 0.74% and so the spectrum does not show clear evidence of this isotope. The combination band transition from the ground vibrational state to $14^00$ is also accessible in the infrared window near 2935 – 2940 cm$^{-1}$ but has a significantly smaller absorption cross section and so the experimental spectrum does not show clear evidence of this combination band either. The line fitting algorithm in PGOPHER was used to fit the $10^01$ spectroscopic constants for OC$^{32}$S (Table 2) and OC$^{34}$S. Line lists and further information is included in the supplementary material. Fayt and coworkers performed global fitting to derive spectroscopic parameters, the results of which are used in the HITRAN database.[74-76] As can be seen in Table 2, the $10^01$ combination band constants derived in this work agree very well with the global fitting results from Fayt and coworkers, as well as the previous microwave spectroscopy results recorded under high temperatures. In addition, the observed line positions agree fairly well with a previously reported experimental linelist,[77] although less agreement was found at higher wavenumbers.

**Table 2.** Spectroscopic constants for OC$^{32}$S found in this work for the $10^01$ combination band and reported previously for both the ground state and the $10^01$ combination band.

| | Ground state (Fayt et al.)[a] | (this work) | Fayt et al.[a), b)] | Bogey, Bauer[78] | Yamada, Klebsch[79] |
|---|---|---|---|---|---|
| Origin | -- | 2918.10510(2) | 2918.1055(2) | -- | -- |
| B | 0.2028567417(5) | 0.2011031(2) | 0.2011033(2) | 0.2011034(5) | 0.20110307(5) |
| D | 4.3412(2) × 10$^{-8}$ | 5.09(6) × 10$^{-8}$ | 5.1509(7) × 10$^{-8}$ | 5.4(8) × 10$^{-8}$ | 4.989(6) × 10$^{-8}$ |
| H | -2.6(3) × 10$^{-15}$ | 3(4) × 10$^{-13}$ | 9.822(6) × 10$^{-13}$ | -- | 3.2(1) × 10$^{-13}$ |

[a)]The error bars were reported to a larger number of digits,[75] which has been reduced to last significant digit in the above table.
[b)]Fayt et al. included even higher order centrifugal distortion constants (L though P) but were vanishingly small and not included in the fit here.[75]

Using contour fitting withing PGOPHER, the rotational temperature was fit to 49.5(1) K, which is approximately 17 K higher than the temperature derived from Pitot impact pressure measurements, at 32(3) K. An alternative way of measuring the temperature is using a standard Boltzmann plot, shown in the supplementary material, which resulted in a temperature of 48.9 K, agreeing very with the PGOPHER contour fit. Yet another way of estimating the temperature is through the analysis of the linewidth. The average Gaussian full-width-at-half-maximum (FWHM) for the OC$^{32}$S peaks was found to be 0.00223(1) cm$^{-1}$. The expected Doppler broadening at 49.5 K is 0.00228 cm$^{-1}$ for OC$^{32}$S at the band origin, which is higher than what is observed. However, Doppler broadening at 32 K, the temperature measured by Pitot impact pressure studies, would results in linewidths of only 0.00183 cm$^{-1}$. Additional influence on the observed linewidth could be from collisional broadening. Previously reported collisional broadening coefficients are on the order of 6 MHz torr$^{-1}$ (0.00015 cm$^{-1}$ mbar$^{-1}$).[80] Although this value is for other collision partners (N$_2$, O$_2$, and self-broadening) and for a different vibrational band, it serves as a good estimation of an expected contribution of 0.00003 cm$^{-1}$ at the 0.2 mbar pressure within the USF. Even taking collisional broadening into account, the resulting Doppler temperature is still 46 K, which is 14 K warmer than the Pitot impact pressure measurements. The boundary layer of the USF is likely to have some influence on the slightly warmer derived temperature. However, there was no systematic trend in the contour fitting residuals indicating that multiple temperatures (for core and boundary layer) would be necessary to fit the data. In addition, the Boltzmann plot (Figure S13 in the supplementary material) is linear when plotting the natural log of the population versus energy level, again agreeing well with the spectrum being well-characterized by a single temperature. It is therefore unclear why there is a discrepancy between the molecular rotational temperature of a stable species within the USF and the Pitot impact pressure measurements, and a further detailed investigation would be desirable. It should be noted that this same nozzle is used in a second CRESU apparatus,[42] and an independent measurement of the flow temperature indicates a very similar temperature of 31(2) K measured by Pitot impact pressure measurements with a different Pitot tube. Insight into some of the underlying assumptions in Pitot measurements and potential causes for discrepancies is part of a recent collaborative investigation.[65]

There are additional peaks at lower wavenumbers in the experimental spectrum, approximately 10 – 15 peaks in total, an example of which is shown panel B of Figure 4. The peaks did not match peak positions for any OCS isotopologue and did not overlap with any hot bands or other combination bands expected to be present at this cold USF temperature, according to HITRAN simulations. It is most likely that these peaks are due to the OCS-Ar van der Waals complex or perhaps the OCS dimer. The OCS-Ar complex has been studied previously and was calculated be bound by approximately 190 cm$^{-1}$.[81-86] The OCS dimer is also bound by similar energies, but dependent on the orientation of the two OCS moieties.[87-91] These are both reasonable candidates for formation at the cold temperatures achieved in the USF. The observed progression has peak spacings of 0.44 cm$^{-1}$, which is very similar to the OC$^{32}$S peak progression. The peaks also have a noticeably wider FWHM than the main progression: approximately 0.0035 cm$^{-1}$ instead of 0.00223 cm$^{-1}$. These attributes, particularly the similarity in rotational constant and thus peak spacing, point towards the unassigned peaks arising from the OCS-Ar complex. However, the peaks extended past the wavenumber region collected during this experiment. A detailed study of these peaks and characterization of the full spectrum is beyond the scope of this work. It should be noted that the concentration of OCS in the USF for this spectroscopic study was below the observed curvature in the second order plot discussed in Section D, which usually indicates excess reagent cluster formation that can interfere with derived second order reaction rate coefficients. The ability to observe these spectroscopic peaks showcases multiple advantages of using DFCS with the USF: first, that complexes that could potentially interfere with kinetic measurements can be observed and perhaps quantified, and second, that thermalized complexes under controlled temperature conditions have the potential to be spectroscopically characterized.

## D. CH + OCS Temperature-dependent Reaction Rate Coefficients

In order to measure reaction rate coefficients for any reaction, it is important to know the limit on the minimum value of the reaction rate coefficient that can be measured, determined by the maximum kinetic time for each Laval nozzle. In our system, the maximum kinetic time for the decay in CH concentration with respect to the delay time between the photolysis and probe laser is defined by the distance from the exit of the Laval nozzle to the fluorescence detection axis. For Nozzle 9, with a measured Mach number of 4.9(2) using an argon buffer gas at a nozzle distance of 26 cm, the gas flow velocity is 523 ms$^{-1}$, corresponding to a maximum reaction time of ~498 μs. Depending on the quality (signal-to-noise ratio) of the data, approximately one half-life of the exponential decay should be measured in order to determine an accurate rate coefficient. In our system, one half-life within the maximum reaction time corresponds to a pseudo-first order rate coefficient of ~1,390 s$^{-1}$. The maximum density of OCS that can be used for the nozzle in question is ~4.3 × 10$^{14}$ molecule cm$^{-3}$, which keeps the maximum mixing ratio of OCS contained within the flow at or below ~1% that of the Ar density. Combining these gives a minimum reaction rate coefficient of ~3.2 × 10$^{-12}$ cm$^3$ molecule$^{-1}$ s$^{-1}$ that can be measured in the available kinetic time for Nozzle 9. Experimentally, the sensitivity limit is controlled by the ability to discriminate the loss of CH due to reaction with OCS from other loss processes discussed later in this section, which is related to the signal-to-noise ratio of the CH signal with no OCS present in the flow.

To measure the temperature-dependent reaction rate coefficients for the loss of CH due to reaction with OCS, the pseudo-first-order loss rates were first measured with varying densities of OCS for a given temperature. The relative CH concentration was monitored by integrating the CH

LIF signal over an acquisition window, set on the oscilloscope to 1.5 μs and re-analysed in post-processing, while randomly varying the time delay between the photolysis and probe laser pulses for each gas pulse from -20 – 300 μs. Negative time delays represent a pre-photolysis regime where no fluorescence is expected to be observed as no CH should be present in the USF. At 0 μs, the photolysis and probe laser pulses are overlapped and so the CH signal is that resulting from CH generated on the fluorescence detection axis. The nozzle position for the experiments was such that the fluorescence signal observed at the longest time delay of 300 μs is that of nascent CH generated at the exit of the Laval nozzle and represents the time taken for the CH radical to reach the fluorescence detection axis. Each kinetic decay consists of 160 datapoints, giving a 2 μs spacing between the laser delay times. A total of five successive scans were taken and averaged together in order to reduce the signal-to-noise on the kinetic decay trace while also not requiring a significant amount of time for data acquisition. No significant difference was observed in the kinetic decay profiles by using a higher number of averages. For each laser delay time, the averaged fluorescence trace was saved and post-processed using an in-house Python analysis script to retrieve the integrated LIF signal at each laser delay time. Further details on this can be found in Section SVI of the supplementary material.

Following analysis of the individual fluorescence traces, the integrated fluorescence signal as a function of the laser delay time was plotted to calculate the pseudo-first-order rate coefficients for each OCS density. The kinetic decay profiles were baseline corrected by averaging the first 14 μs at negative laser delay times and subtracting the value from the rest of the kinetic decay trace. There is a sharp rise in the integrated fluorescence signal on short timescales (≤ 10 μs) relating to the fast relaxation of rotationally and/or vibrationally excited CH radicals to the rovibrational level of the X $^2\Pi$ ground state probed via LIF. In addition, the PMT gate can interfere with early times of the fluorescence signal. Data were fit from several times between 20 – 36 μs and the CH loss rates due to reaction with OCS were found to be independent of the fitting window, suggesting the single exponential decay fitting parameter is free from CH rovibrational relaxation or PMT gate interference after 20 μs. Thus, a single exponential decay function given by Equation 5 was fit to the integrated fluorescence signal for timescales longer than 30 μs. Examples of the temporal evolution of the integrated fluorescence signal for four OCS densities are displayed in Figure 5. As expected, an exponential decay in the integrated fluorescence signal is observed in the presence of OCS, in which the pseudo-first-order rate coefficient increases from 3,071 s$^{-1}$ to 46,101 s$^{-1}$ for OCS densities of 0.00 molecule cm$^{-3}$ and 1.54 × 10$^{14}$ molecule cm$^{-3}$, respectively. Further discussion on this is given later in this section when discussing the retrieval of the reaction rate coefficient.

$$I_{LIF}(t) = Ae^{-k't} \quad (5)$$

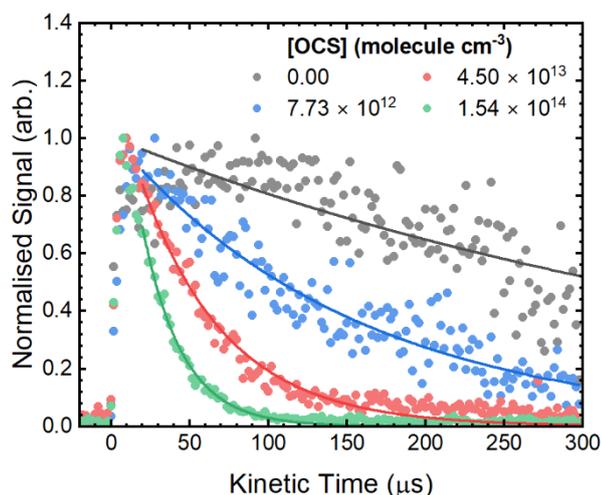

**Figure 5.** Representative normalised transient CH integrated fluorescence signal as a function of laser delay time taken at 32(3) K and a total density of 4(1) × 10$^{16}$ molecule cm$^{-3}$, together with single exponential decay fits for [OCS] = 0.00 (black), 7.73 × 10$^{12}$ (blue), 4.50 × 10$^{13}$ (red), 1.54 × 10$^{14}$ (green) molecule cm$^{-3}$, respectively.

The pseudo-first-order rate coefficients, $k'$, are derived at known OCS densities by fitting Equation 5 to the kinetic decay profiles shown in Figure 5. For each temperature, reaction rate coefficients were then obtained from the variation of the pseudo-first-order rate coefficient with OCS density using Equation 6. Here, $k_{CH+OCS}$ represents the bimolecular reaction rate coefficient for reaction with OCS and $k_0$ represents a loss rate of CH due to diffusion out of the detection zone, reaction with precursor, and reaction with other impurities (such as photolytic products, or 2-methyl-2-butene present in the CHBr$_3$ sample at 60-120 ppm as stabiliser).

$$k' = k_{CH+OCS}[OCS] + k_0 \qquad (6)$$

A representative plot of the pseudo-first-order rate coefficient, $k'$, as a function of OCS number density is presented in Figure 6 for Nozzle 9 using Ar as the buffer gas. As highlighted in Figure 6A, the pseudo-first-order rate coefficient for a given temperature increased linearly with respect to the OCS density, until OCS densities were greater than 1 × 10$^{14}$ molecule cm$^{-3}$. Above this OCS density, curvature was observed in the plot where the pseudo-first-order rate coefficient appeared to remain approximately constant with increasing OCS density. This curavature is likely due to a loss of OCS in which OCS dimers, higher order oligomers, or perhaps clusters with Ar or the photolytic precursor are formed, reducing the OCS monomer density from that expected. There is further evidence for OCS clusters in the infrared spectrum shown in Figure 4 and discussed above. Hence only the linear portion of the plot is used to derive the reaction rate coefficient using a linear least squares regression fit where the reaction rate coefficient is retrieved from the slope of the fit at a given temperature. The average reaction rate coefficient was then derived to be 3.9(4) × 10$^{-10}$ cm$^3$ molecule$^{-1}$ s$^{-1}$, with an intercept value of 4700(900) s$^{-1}$ at 32(3) K and an average total density of 4(1) × 10$^{16}$ molecule cm$^{-3}$. Propagation of error is discussed extensively in Section SVI of the supplementary material. Measurements of the reaction rate coefficient were repeated at 32(3) K a total of 12 times to verify the experimental reproducibility. In addition, the reaction rate coefficient was found to be independent of the precursor concentration and the rovibronic transition being probed, as shown in Figure 6B below.

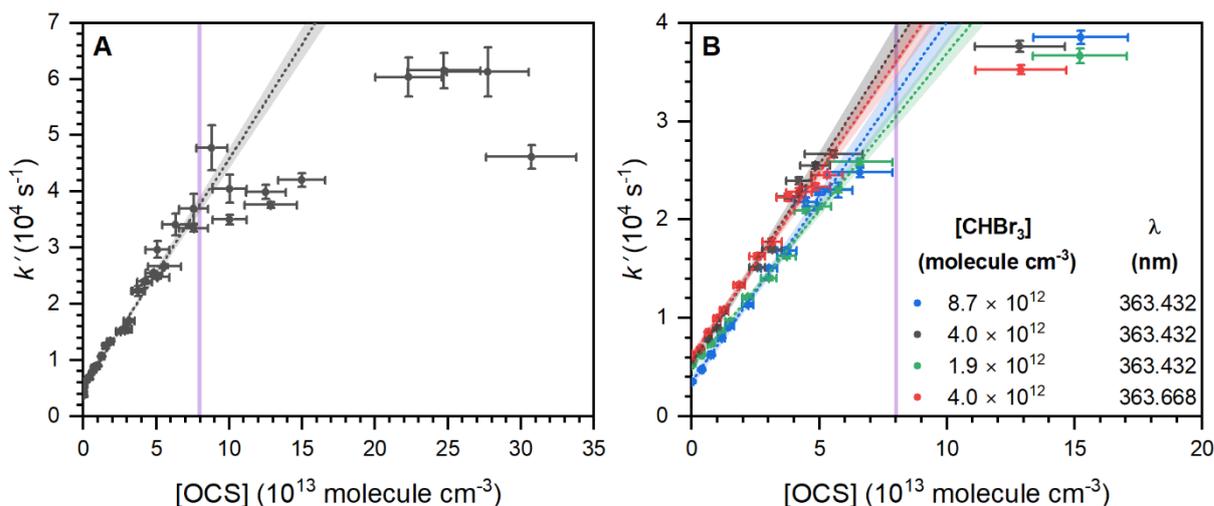

**Figure 6.** Loss rate of CH as a function of OCS density at 32(3) K, a total density of $4(1) \times 10^{16}$ molecule cm$^{-3}$ and [CHBr$_3$] = $4.5 \times 10^{12}$ molecule cm$^{-3}$ with linear fits to determine $k$(32(3) K) and 95% confidence bands. A: Study to determine the [OCS] concentration at which curvature is observed. Data where [OCS] > $8 \times 10^{13}$ molecule cm$^{-3}$ were excluded from the fit as indicated by the purple line. $k$(32(3) K) = $4.0(1) \times 10^{-10}$ cm$^3$ molecule$^{-1}$ s$^{-1}$. B: Data for $8.7 \times 10^{12}$ molecule cm$^{-3}$ (blue), [CHBr$_3$] = $4.0 \times 10^{12}$ molecule cm$^{-3}$ (black), and $1.9 \times 10^{12}$ molecule cm$^{-3}$ (green), using the Q$_2$(1) transition at 363.432 nm. The second [CHBr$_3$] used to examine the CH propensity using the Q$_1$(1) transition at 363.668 nm (red). $k$(32(3) K) = $3.7(1) \times 10^{-10}$, $4.0(2) \times 10^{-10}$, $3.2(1) \times 10^{-10}$, and $3.7(1) \times 10^{-10}$ cm$^3$ molecule$^{-1}$ s$^{-1}$ for the blue, black, green, and red data, respectively.

The temperature dependence of the reaction rate coefficient was then examined by changing the Laval nozzle and thus, the USF temperature. The measured reaction rate coefficients and experimental conditions can be found in Table S5 of the supplementary material. The average values of the reaction rate coefficient are shown as a function of temperature in purple in Figure 7. Shown in Figure 7A is an enlarged plot showing the measured $k(T)$ values in this work, as well as those from previous experiments and theoretical calculations for temperatures less than 500 K.[92,93] The experimental errors in this work are reported as one standard deviation from the average value. These values of $k(T)$ are then compared to those calculated at the collision limit using collision capture theory (CCT), discussed in a previous publication,[93] for temperatures less than 3000 K in Figure 7B.

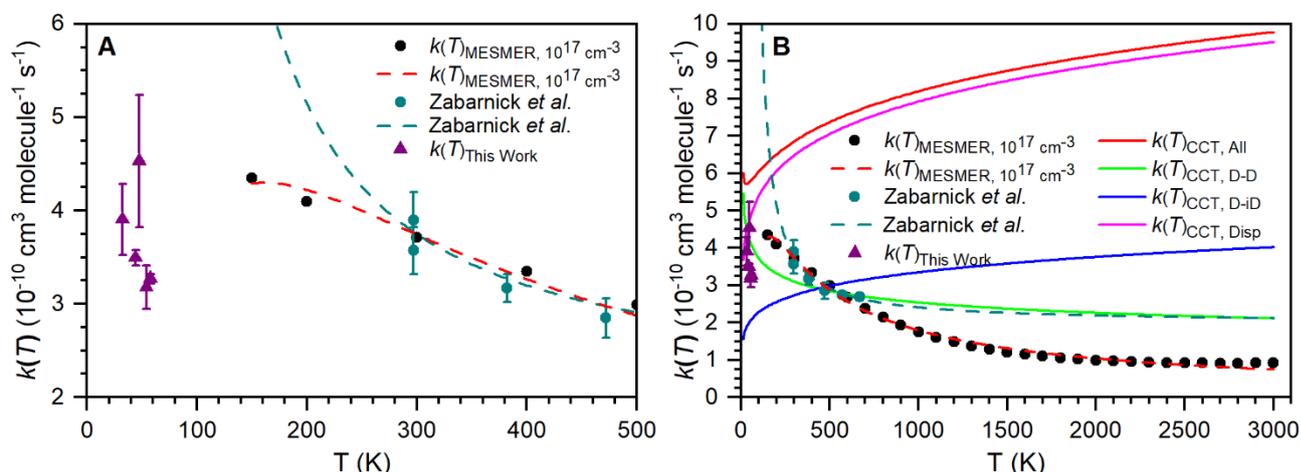

**Figure 7.** Averaged values of $k(T)$ for CH + OCS as a function of temperature measured in this work (purple), compared to previous theoretical calculations (black) and experimental values (teal). The errors represent 1 standard deviation of the average value. A: Zoom in to clearly show the measured values of $k(T)$ in this work. B: Comparison of all data with collision capture theory, with the overall rate coefficient shown in red and the individual contributions from dipole – dipole, dipole – induced dipole, and dispersion forces shown in green, blue, and pink, respectively. For details of these calculations, see a previous publication.[93]

As can be seen in Figure 7A, the rate coefficient showed no significant temperature dependence over the range of 32(3) < T < 58(5) K. The pressure dependence of this reaction was not examined here, but the reaction is expected to be pressure independent based on prior theoretical work.[93] Also shown in Figure 7A and 7B are the measurements made by Zabarnick et al.[92] in the range of 297 < T < 667 K using a flow tube type experiment. The authors reported a slight negative temperature dependence, which is also supported by rate coefficients calculated from statistical rate theory. The average value of $k(32(3)$ K$)$ was measured to be $3.9(4) \times 10^{-10}$ cm$^3$ molecule$^{-1}$ s$^{-1}$ which supports the suggestion of a negative temperature dependence as the temperature decreases of the modified Arrhenius fit to the theoretical data around 150 K. However, a positive or negative temperature dependence cannot be confirmed from the measurements made in this work and further experimental measurements are required from 58 K to 150 K in order to fully examine this temperature dependence.

It should be noted that using a nitrogen buffer gas to achieve the higher USF temperatures when measuring the CH + OCS reaction is difficult due to the competitive reaction of CH + N$_2$ when N$_2$ is in large excess as the USF buffer gas, and the temperature and pressure dependence of the CH + N$_2$ second order rate coefficient. The expected second order rate coefficient can be determined using the previously presented results from global fitting.[94, 95]

$$k_0(T, [\text{M} = \text{Ar}]) = (1.6 \times 10^{-31}) \times \left(\frac{T}{298}\right)^{-2.2} \times [\text{Ar}] \text{ cm}^3 \text{ molecule}^{-1} \text{ s}^{-1} \qquad (7)$$

It was observed[94, 95] that the third body identity does not impact the rate expression given by Equation 7. For the nozzle Print 2, for example, operating at 89 K and a total density of $12(3) \times 10^{16}$ molecule cm$^{-3}$, the CH + N$_2$ reaction rate coefficient is $2.7 \times 10^{-13}$ cm$^3$ molecule$^{-1}$ s$^{-1}$, which corresponds to a $k'$ value of ~$3.1 \times 10^4$ s$^{-1}$ with no OCS present in the flow. This value is already approximately the same as the $k'$ value expected for some of the highest OCS concentrations reported here. We would then be attempting to observe relatively small changes due to changing OCS concentrations on top of an already fast kinetic decay. We are currently working on designing additional nozzles that operate at lower total densities while maintaining higher temperatures when $N_2$ is used as a buffer gas to continue studies of CH radical reactions across a wider range of temperatures.

## IV. Conclusion

We presented the development of an advanced research tool, HILTRAC, designed to collect information on the reaction kinetics and dynamics of important astrochemical reactions. A well-established kinetic detection technique, PLP-LIF, has been successfully coupled to the USF, and the first implementation of DFCS coupled to a USF has been presented in this work. The low temperature environment required to measure important astrochemical reactions has been generated using the pulsed CRESU technique, which has been characterised in this work by pressure impact measurements, with the achievable temperature and total gas density ranges being 32(3) – 111(9) K and 4(1) – 14(3) $\times 10^{16}$ molecule cm$^{-3}$, respectively, when utilising both Ar and $N_2$ as a buffer gas with five different Laval nozzles. With the current machined or 3D printed Laval nozzles in use, we can achieve between 239 and 498 µs of hydrodynamic time. Also presented in this work is the capability to fully examine the USF in three dimensions with fully automated collection of the impact pressure as a function of the axial, radial, and vertical distances, allowing for detailed visualisation of the USF. Further investigation of the USF flow temperature has been provided by fitting the LIF spectrum of CH and DFCS spectrum of OCS, where the USF temperature was found to be ~10 – 15 K greater than the temperature obtained from pressure impact measurements. The DCFS technique was further used to measure accurate spectroscopic constants and report linelists for OC$^{32}$S and OC$^{34}$S for the $10^01$ combination band. Temperature dependent reaction rate coefficients for the CH + OCS reaction have been measured in the temperature range of 32(3) – 58(5) K, measuring $k(T)$ at 32(3) K to be
3.9(4) $\times 10^{-10}$ cm$^3$ molecule$^{-1}$ s$^{-1}$. No observable temperature dependence was found over the temperature range used in this work, with the reaction rate coefficients measured showing no dependence on the radical precursor concentration or the rovibronic transition used. Extensive uncertainty analysis for all aspects of the work presented has also been performed to fully characterise the HILTRAC apparatus. While product branching fractions have not been measured in this work, the successful implementation of DFCS offers real prospects of observing multiple molecular species involved in a reaction in future studies. In addition, the HILTRAC apparatus has been designed to make future modification possible, such as coupling other detection methods to the USF, making this highly versatile tool for studying the chemistry of planetary atmospheres or the interstellar medium.

## Supplementary Material

Additional information on the HILTRAC apparatus characterization, including uncertainty analysis, calibrations, and frequency comb spectral analysis, can be found in the supplementary material. Also included is further detail on the CH + OCS kinetic measurements and OCS spectroscopic parameters and linelists.

# Acknowledgements

This project has received funding from the European Research Council (ERC) under the European Union's Horizon 2020 research and innovation programme (grant agreement No. 948525). The authors would like to thank Dr Niclas West, Dr Mark Blitz, and Dr Kevin Douglas at the University of Leeds for useful discussions during the commissioning of HILTRAC. The authors would also like to thank University of Birmingham students Theobold Stevens and Thomas Edwards for contributing to data collection and nozzle design.

# Author Declarations

The authors declare no competing financial interests.

# Data Availability

Most data are already available in the supplementary material. If additional data are required, please contact the corresponding author.

# References

(1) Herbst, E.; van Dishoeck, E. F. Complex Organic Interstellar Molecules. In *Annual Review of Astronomy and Astrophysics, Vol 47*, Blandford, R., Kormendy, J., VanDishoeck, E. Eds.; Annual Review of Astronomy and Astrophysics, Vol. 47; Annual Reviews, 2009; pp 427-480.
(2) McGuire, B. A. 2021 Census of Interstellar, Circumstellar, Extragalactic, Protoplanetary Disk, and Exoplanetary Molecules. *Astrophysical Journal Supplement Series* **2022**, *259* (2), 51. DOI: 10.3847/1538-4365/ac2a48.
(3) Swings, P.; Rosenfeld, L. Considerations Regarding Interstellar Molecules. *Astrophys. J.* **1937**, *86* (4), 483-486. DOI: 10.1086/143880.
(4) Motiyenko, R. A.; Belloche, A.; Garrod, R. T.; Margulès, L.; Müller, H. S. P.; Menten, K. M.; Guillemin, J.-C. Millimeter- and Submillimeter-wave Spectroscopy of Thioformamide and Interstellar Search Toward Sgr B2(N)*. *Astronomy and Astrophysics* **2020**, *642*, A29.
(5) Rodríguez-Almeida, L. F.; Jiménez-Serra, I.; Rivilla, V. M.; Martín-Pintado, J.; Zeng, S.; Tercero, B.; de Vicente, P.; Colzi, L.; Rico-Villas, F.; Martín, S.; Requena-Torres, M. A. Thiols in the Interstellar Medium: First Detection of HC(O)SH and Confirmation of $C_2H_5SH$. *The Astrophysical Journal Letters* **2021**, *912* (1), L11. DOI: 10.3847/2041-8213/abf7cb.
(6) Sanz-Novo, M.; Rivilla, V. M.; Müller, H. S. P.; Jiménez-Serra, I.; Martín-Pintado, J.; Colzi, L.; Zeng, S.; Megías, A.; López-Gallifa, Á.; Martínez-Henares, A.; Tercero, B.; de Vicente, P.; San Andrés, D.; Martín, S.; Requena-Torres, M. A. Discovery of Thionylimide, HNSO, in Space: The First N-, S-, and O-bearing Interstellar Molecule. *The Astrophysical Journal Letters* **2024**, *965* (2), L26. DOI: 10.3847/2041-8213/ad3945.
(7) ALMA Observatory. *Atacama Large Millimeter/submillimeter Array*. 2021. https://www.almaobservatory.org/en/home/ (accessed 15th May 2024).
(8) Space Telescope Science Institute. *Webb Space Telescope*. 2017. https://webbtelescope.org/home (accessed 15th May 2024).
(9) Millar, T. J.; Walsh, C.; Van de Sande, M.; Markwick, A. J. The UMIST Database for Astrochemistry 2022. *Astronomy and Astrophysics* **2024**, *682*, 10.1051/0004-6361/202346908.
(10) Cernicharo, J.; Cabezas, C.; Agúndez, M.; Tercero, B.; Pardo, J. R.; Marcelino, N.; Gallego, J. D.; Tercero, F.; López-Pérez, J. A.; de Vicente, P. TMC-1, the Starless Core Sulfur Factory: Discovery of NCS, HCCS, $H_2CCS$, $H_2CCCS$, and $C_4S$ and detection of $C_5S$*. *Astronomy and Astrophysics* **2021**, *648*, 10.1051/0004-6361/202140642.

(11) Vidal, T. H. G.; Loison, J. C.; Jaziri, A. Y.; Ruaud, M.; Gratier, P.; Wakelam, V. On the Reservoir of Sulphur in Dark Clouds: Chemistry and Elemental Abundance Reconciled. *Monthly Notices of the Royal Astronomical Society* **2017**, *469* (1), 435-447. DOI: 10.1093/mnras/stx828.
(12) McElroy, D.; Walsh, C.; Markwick, A. J.; Cordiner, M. A.; Smith, K.; Millar, T. J. The UMIST Database for Astrochemistry 2012. *Astronomy & Astrophysics* **2013**, *550*, article no: A36 [no pagination], Article. DOI: 10.1051/0004-6361/201220465.
(13) Martínez, E.; Albaladejo, J.; Jiménez, E.; Notario, A.; Aranda, A. Kinetics of the Reaction of $CH_3S$ with $NO_2$ as a Function of Temperature. *Chemical Physics Letters* **1999**, *308* (1), 37-44. DOI: https://doi.org/10.1016/S0009-2614(99)00579-5.
(14) Smith, I. W. M.; Barnes, P. W. Advances in Low Temperature Gas-Phase Kinetics. *Annual Reports Section "C" (Physical Chemistry)* **2013**, *109* (0), 140-166, 10.1039/C3PC90011H. DOI: 10.1039/C3PC90011H.
(15) Sims, I. R.; Smith, I. W. M. Rate Constants for the Radical-Radical Reaction Between CN and $O_2$ at Temperatures Down to 99 K. *Chemical Physics Letters* **1988**, *151* (6), 481-484. DOI: https://doi.org/10.1016/S0009-2614(88)85021-8.
(16) Hawley, M.; Mazely, T. L.; Randeniya, L. K.; Smith, R. S.; Zeng, X. K.; Smith, M. A. A Free Jet Flow Reactor for Ion/Molecule Reaction Studies at Very Low Energies. *International Journal of Mass Spectrometry and Ion Processes* **1990**, *97* (1), 55-86. DOI: https://doi.org/10.1016/0168-1176(90)85040-9.
(17) Rowe, B. R.; Dupeyrat, G.; Marquette, J. B.; Gaucherel, P. Study of the Reactions $N_2^+ + 2N_2 \rightarrow N_4^+ + N_2$ and $O_2^+ + 2O_2 \rightarrow O_4^+ + O_2$ from 20 to 160 K by the CRESU Technique. *Journal of Chemical Physics* **1984**, *80* (10), 4915-4921, Article. DOI: 10.1063/1.446513.
(18) Rowe, B. R.; Marquette, J. B. Cresu Studies of Ion Molecule Reactions. *International Journal of Mass Spectrometry* **1987**, *80*, 239-254, Article. DOI: 10.1016/0168-1176(87)87033-7.
(19) Rowe, B.; Canosa, A.; Heard, D. E. Uniform Supersonic Flows in Chemical Physics. WORLD SCIENTIFIC (EUROPE): 2021; p 728.
(20) Cooke, I. R.; Sims, I. R. Experimental Studies of Gas-Phase Reactivity in Relation to Complex Organic Molecules in Star-Forming Regions. *Acs Earth and Space Chemistry* **2019**, *3* (7), 1109-1134. DOI: 10.1021/acsearthspacechem.9b00064.
(21) Potapov, A.; Canosa, A.; Jimenez, E.; Rowe, B. Uniform Supersonic Chemical Reactors: 30 Years of Astrochemical History and Future Challenges. *Angewandte Chemie-International Edition* **2017**, *56* (30), 8618-8640, Review. DOI: 10.1002/anie.201611240.
(22) Heard, D. E. Rapid Acceleration of Hydrogen Atom Abstraction Reactions of OH at Very Low Temperatures through Weakly Bound Complexes and Tunneling. *Accounts of Chemical Research* **2018**, *51* (11), 2620-2627. DOI: 10.1021/acs.accounts.8b00304.
(23) Rowe, B. R.; Marquette, J. B.; Dupeyrat, G.; Ferguson, E. E. Reactions of $He^+$ and $N^+$ Ions with Several Molecules at 8 K. *Chemical Physics Letters* **1985**, *113* (4), 403-406. DOI: https://doi.org/10.1016/0009-2614(85)80391-2.
(24) Sims, I. R.; Queffelec, J. L.; Defrance, A.; Rebrionrowe, C.; Travers, D.; Rowe, B. R.; Smith, I. W. M. Ultra-Low Temperature Kinetics of Neutral-Neutral Reactions - The Reaction $CN+O_2$ down to 26 K. *Journal of Chemical Physics* **1992**, *97* (11), 8798-8800, Letter. DOI: 10.1063/1.463349.
(25) Sims, I. R.; Queffelec, J. L.; Defrance, A.; Rebrionrowe, C.; Travers, D.; Bocherel, P.; Rowe, B. R.; Smith, I. W. M. Ultralow Temperature Kinetics of Neutral-Neutral Reactions. The Technique and Results for the Reactions $CN+O_2$ down to 13 K and $CN+NH_3$ down to 25 K. *Journal of Chemical Physics* **1994**, *100* (6), 4229-4241, Article. DOI: 10.1063/1.467227.
(26) James, P. L.; Sims, I. R.; Smith, I. W. M. Total and State-To-State Rate Coefficients for Rotational Energy Transfer in Collisions between $NO(X\ ^2\Pi)$ and He at Temperatures down to 15 K. *Chemical Physics Letters* **1997**, *272* (5), 412-418. DOI: https://doi.org/10.1016/S0009-2614(97)00539-3.

(27) Daugey, N.; Caubet, P.; Retail, B.; Costes, M.; Bergeat, A.; Dorthe, G. Kinetic Measurements on Methylidyne Radical Reactions with Several Hydrocarbons at Low Temperatures. *Physical Chemistry Chemical Physics* **2005**, *7* (15), 2921-2927, 10.1039/B506096F. DOI: 10.1039/B506096F.
(28) Daugey, N.; Caubet, P.; Bergeat, A.; Costes, M.; Hickson, K. M. Reaction Kinetics to Low Temperatures. Dicarbon + Acetylene, Methylacetylene, Allene and Propene from 77 ≤ T ≤ 296 K. *Physical Chemistry Chemical Physics* **2008**, *10* (5), 729-737, 10.1039/B710796J. DOI: 10.1039/B710796J.
(29) Atkinson, D. B.; Smith, M. A. Design and Characterization of Pulsed Uniform Supersonic Expansions for Chemical Applications. *Review of Scientific Instruments* **1995**, *66* (9), 4434-4446, Article. DOI: 10.1063/1.1145338.
(30) Lee, S.; Hoobler, R. J.; Leone, S. R. A Pulsed Laval Nozzle Apparatus with Laser Ionization Mass Spectroscopy for Direct Measurements of Rate Coefficients at Low Temperatures with Condensable Gases. *Review of Scientific Instruments* **2000**, *71* (4), 1816-1823, Article. DOI: 10.1063/1.1150542.
(31) Spangenberg, T.; Köhler, S.; Hansmann, B.; Wachsmuth, U.; Abel, B.; Smith, M. A. Low-Temperature Reactions of OH Radicals with Propene and Isoprene in Pulsed Laval Nozzle Expansions. *The Journal of Physical Chemistry A* **2004**, *108* (37), 7527-7534. DOI: 10.1021/jp031228m.
(32) Taylor, S. E.; Goddard, A.; Blitz, M. A.; Cleary, P. A.; Heard, D. E. Pulsed Laval Nozzle Study of the Kinetics of OH with Unsaturated Hydrocarbons at Very Low Temperatures. *Physical Chemistry Chemical Physics* **2008**, *10* (3), 422-437, Article. DOI: 10.1039/b711411g.
(33) Oldham, J. M.; Abeysekera, C.; Joalland, B.; Zack, L. N.; Prozument, K.; Sims, I. R.; Park, G. B.; Field, R. W.; Suits, A. G. A Chirped-Pulse Fourier-Transform Microwave/Pulsed Uniform Flow Spectrometer. I. The Low-Temperature Flow System. *Journal of Chemical Physics* **2014**, *141* (15), 7, Article. DOI: 10.1063/1.4897979.
(34) Jimenez, E.; Ballesteros, B.; Canosa, A.; Townsend, T. M.; Maigler, F. J.; Napal, V.; Rowe, B. R.; Albaladejo, J. Development of a Pulsed Uniform Supersonic Gas Expansion System based on an Aerodynamic Chopper for Gas Phase Reaction Kinetic Studies at Ultra-Low Temperatures. *Review of Scientific Instruments* **2015**, *86* (4), 17, Article. DOI: 10.1063/1.4918529.
(35) Morales, S. Le Hacheur Aérodynamique un Nouvel Instrument Dédié aux Processus Réactionnels à Ultra-Basse Température. Université de Rennes, 2009.
(36) Cheikh-Sid-Ely, S.; Morales, S. B.; Guillemin, J. C.; Klippenstein, S. J.; Sims, I. R. Low Temperature Rate Coefficients for the Reaction CN + HC$_3$N. *Journal of Physical Chemistry A* **2013**, *117* (46), 12155-12164, Article. DOI: 10.1021/jp406842q.
(37) Rowe, B.; Morales, S. B. Aerodynamic Chopper for Gas Flow Pulsing. WO2011018571, 2009.
(38) Tango, W. J.; Link, J. K.; Zare, R. N. Spectroscopy of K$_2$ Using Laser‐Induced Fluorescence. *The Journal of Chemical Physics* **1968**, *49* (10), 4264-4268. DOI: 10.1063/1.1669869 (acccessed 5/2/2024).
(39) Jimenez, E.; Antinolo, M.; Ballesteros, B.; Canosa, A.; Albaladejo, J. First Evidence of the Dramatic Enhancement of the Reactivity of Methyl Formate (HC(O)OCH$_3$) with OH at Temperatures of the Interstellar Medium: a Gas-Phase Kinetic Study between 22 K and 64 K. *Physical Chemistry Chemical Physics* **2016**, *18* (3), 2183-2191, Article. DOI: 10.1039/c5cp06369h.
(40) Shannon, R. J.; Blitz, M. A.; Goddard, A.; Heard, D. E. Accelerated Chemistry in the Reaction between the Hydroxyl Radical and Methanol at Interstellar Temperatures Facilitated by Tunnelling. *Nature Chemistry* **2013**, *5* (9), 745-749. DOI: 10.1038/nchem.1692.
(41) Caravan, R. L.; Shannon, R. J.; Lewis, T.; Blitz, M. A.; Heard, D. E. Measurements of Rate Coefficients for Reactions of OH with Ethanol and Propan-2-ol at Very Low Temperatures. *Journal of Physical Chemistry A* **2015**, *119* (28), 7130-7137, Article. DOI: 10.1021/jp505790m.

(42) West, N. A.; Millar, T. J.; Van de Sande, M.; Rutter, E.; Blitz, M. A.; Decin, L.; Heard, D. E. Measurements of Low Temperature Rate Coefficients for the Reaction of CH with $CH_2O$ and Application to Dark Cloud and AGB Stellar Wind Models. *Astrophysical Journal* **2019**, *885* (2), 11, Article. DOI: 10.3847/1538-4357/ab480e.
(43) Gupta, D.; Ely, S. C. S.; Cooke, I. R.; Guillaume, T.; Khedaoui, O. A.; Hearne, T. S.; Hays, B. M.; Sims, I. R. Low Temperature Kinetics of the Reaction Between Methanol and the CN Radical. *Journal of Physical Chemistry A* **2019**, *123* (46), 9995-10003, Article. DOI: 10.1021/acs.jpca.9b08472.
(44) Yin, T.; Ma, L.; Cheng, M.; Gao, H. Observation of the Electronic Band System $2^3\Sigma_g^-–a^3\Pi_u$ of $C_2$ in the Vacuum Ultraviolet Region. *The Journal of Chemical Physics* **2023**, *158* (17), 174304. DOI: 10.1063/5.0149708 (acccessed 5/2/2024).
(45) McKee, K.; Blitz, M. A.; Hughes, K. J.; Pilling, M. J.; Qian, H.-B.; Taylor, A.; Seakins, P. W. H Atom Branching Ratios from the Reactions of CH with $C_2H_2$, $C_2H_4$, $C_2H_6$, and neo-$C_5H_{12}$ at Room Temperature and 25 Torr. *The Journal of Physical Chemistry A* **2003**, *107* (30), 5710-5716. DOI: 10.1021/jp021613w.
(46) Douglas, K. M.; Blitz, M. A.; Feng, W.; Heard, D. E.; Plane, J. M. C.; Rashid, H.; Seakins, P. W. Low Temperature Studies of the Rate Coefficients and Branching Ratios of Reactive Loss vs Quenching for the Reactions of $^1CH_2$ with $C_2H_6$, $C_2H_4$, $C_2H_2$. *Icarus* **2019**, *321*, 752-766. DOI: https://doi.org/10.1016/j.icarus.2018.12.027.
(47) Hickson, K. M.; Loison, J.-C.; Wakelam, V. Temperature Dependent Product Yields for the Spin Forbidden Singlet Channel of the $C(^3P)+C_2H_2$ Reaction. *Chemical Physics Letters* **2016**, *659*, 70-75. DOI: https://doi.org/10.1016/j.cplett.2016.07.004.
(48) Soorkia, S.; Trevitt, A. J.; Selby, T. M.; Osborn, D. L.; Taatjes, C. A.; Wilson, K. R.; Leone, S. R. Reaction of the $C_2H$ Radical with 1-Butyne ($C_4H_6$): Low-Temperature Kinetics and Isomer-Specific Product Detection. *Journal of Physical Chemistry A* **2010**, *114* (9), 3340-3354.
(49) Suas-David, N.; Thawoos, S.; Suits, A. G. A Uniform Flow-Cavity Ring-Down Spectrometer (UF-CRDS): A New Setup for Spectroscopy and Kinetics at Low Temperature. *Journal of Chemical Physics* **2019**, *151* (24), 11, Article. DOI: 10.1063/1.5125574.
(50) Abeysekera, C.; Zack, L. N.; Park, G. B.; Joalland, B.; Oldham, J. M.; Prozument, K.; Ariyasingha, N. M.; Sims, I. R.; Field, R. W.; Suits, A. G. A Chirped-Pulse Fourier-Transform Microwave/Pulsed Uniform Flow Spectrometer. II. Performance and Applications for Reaction Dynamics. *Journal of Chemical Physics* **2014**, *141* (21), 9, Article. DOI: 10.1063/1.4903253.
(51) Hays, B. M.; Guillaume, T.; Hearne, T. S.; Cooke, I. R.; Gupta, D.; Abdelkader Khedaoui, O.; Le Picard, S. D.; Sims, I. R. Design and Performance of an E-band Chirped Pulse Spectrometer for Kinetics Applications: OCS – He Pressure Broadening. *Journal of Quantitative Spectroscopy and Radiative Transfer* **2020**, *250*, 107001. DOI: https://doi.org/10.1016/j.jqsrt.2020.107001.
(52) Guillaume, T.; Hays, B. M.; Gupta, D.; Cooke, I. R.; Khedaoui, O. A.; Hearne, T. S.; Drissi, M.; Sims, I. Product-Specific Reaction Kinetics in Continuous Uniform Supersonic Flows probed by Chirped-Pulse Microwave Spectroscopy. *Journal of Chemical Physics* **[Submitted 2024]**.
(53) Lang, J.; Foley, C. D.; Thawoos, S.; Behzadfar, A.; Liu, Y.; Zádor, J.; Suits, A. Reaction Dynamics of $S(^3P)$ with 1,3-Butadiene and Isoprene: Crossed Beam Scattering, Low Temperature Flow Experiments, and High-Level Electronic Structure Calculations. *Faraday Discussions* **2024**, 10.1039/D4FD00009A. DOI: 10.1039/D4FD00009A.
(54) Picqué, N.; Hänsch, T. W. Frequency Comb Spectroscopy. *Nature Photonics* **2019**, *13* (3), 146-157. DOI: 10.1038/s41566-018-0347-5.
(55) Luo, P.-L.; Chen, I. Y.; Khan, M. A. H.; Shallcross, D. E. Direct Gas-Phase Formation of Formic Acid through Reaction of Criegee Intermediates with Formaldehyde. *Communications Chemistry* **2023**, *6* (1), 130. DOI: 10.1038/s42004-023-00933-2.

(56) Bjork, B. J.; Bui, T. Q.; Heckl, O. H.; Changala, P. B.; Spaun, B.; Heu, P.; Follman, D.; Deutsch, C.; Cole, G. D.; Aspelmeyer, M.; Okumura, M.; Ye, J. Direct Frequency Comb Measurement of OD + CO → DOCO Kinetics. *Science* **2016**, *354* (6311), 444-448, Article. DOI: 10.1126/science.aag1862.

(57) Long, D. A.; Cich, M. J.; Mathurin, C.; Heiniger, A. T.; Mathews, G. C.; Frymire, A.; Rieker, G. B. Nanosecond Time-Resolved Dual-Comb Absorption Spectroscopy. *Nature Photonics* **2024**, *18* (2), 127-131. DOI: 10.1038/s41566-023-01316-8.

(58) Dubroeucq, R. Development of a Cavity-Enhanced Fourier Transform Spectrometer based on Optical Frequency Combs for Laboratory Astrophysics. Université de Rennes, 2023. https://theses.hal.science/tel-04523315.

(59) Dudás, E.; Suas-David, N.; Brahmachary, S.; Kulkarni, V.; Benidar, A.; Kassi, S.; Charles, C.; Georges, R. High-Temperature Hypersonic Laval Nozzle for non-LTE Cavity Ringdown Spectroscopy. *The Journal of Chemical Physics* **2020**, *152* (13), 134201. DOI: 10.1063/5.0003886 (acccessed 5/2/2024).

(60) Bonnamy, A.; Georges, R.; Hugo, E.; Signorell, R. IR Signature of $(CO_2)_N$ Clusters: Size, Shape and Structural Effects. *Physical Chemistry Chemical Physics* **2005**, *7* (5), 963-969, 10.1039/B414670K. DOI: 10.1039/B414670K.

(61) Chen, C.; Sheng, Y.; Yu, S.; Ma, X. Investigation of the Collisional Quenching of CH(A $^2\Delta$ and B $^2\Sigma^-$) by Ar, $O_2$, $CS_2$, Alcohol, and Halomethane Molecules. *The Journal of Chemical Physics* **1994**, *101* (7), 5727-5730. DOI: 10.1063/1.467358 (acccessed 5/3/2024).

(62) Cooper, J. L.; Whitehead, J. C. Collisional Removal Rates for Electronically Excited CH Radicals B $^2\Sigma^-$ and C $^2\Sigma^+$. *Journal of the Chemical Society, Faraday Transactions* **1992**, *88* (16), 2323-2327, 10.1039/FT9928802323. DOI: 10.1039/FT9928802323.

(63) Roberts, F. C.; Lewandowski, H. J.; Hobson, B. F.; Lehman, J. H. A Rapid, Spatially Dispersive Frequency Comb Spectrograph Aimed at Gas Phase Chemical Reaction Kinetics. *Molecular Physics* **2020**, *118* (16), 9, Article. DOI: 10.1080/00268976.2020.1733116.

(64) Roberts, F. C. Direct Frequency Comb Spectroscopy: Commissioning a New Apparatus and Investigating the Rovibrational Spectroscopy of $CH_2X_2$ Molecules. University of Leeds, 2022. https://etheses.whiterose.ac.uk/31974/.

(65) Driver, L.; Douglas, K. M.; Lucas, D. I.; Guillaume, T.; Lehman, J. H.; Kapur, N.; Heard, D. E.; de Boer, G. N. Developing a Predictive Model for Low Temperature Laval Nozzles with Application in Chemical Kinetics. *Physics of Fluids* **[Submitted 2024]**.

(66) Hay, S.; Shokoohi, F.; Callister, S.; Wittig, C. Collisional Metastability of High Rotational States of CN(X $^2\Sigma^+$, v″ = 0) *. *Chemical Physics Letters* **1985**, *118* (1), 6-11. DOI: https://doi.org/10.1016/0009-2614(85)85255-6.

(67) Mertens, L. A.; Labiad, H.; Denis-Alpizar, O.; Fournier, M.; Carty, D.; Le Picard, S. D.; Stoecklin, T.; Sims, I. R. Rotational Energy Transfer in Collisions Between CO and Ar at Temperatures from 293 to 30 K. *Chemical Physics Letters* **2017**, *683*, 521-528. DOI: https://doi.org/10.1016/j.cplett.2017.05.052.

(68) Labiad, H.; Fournier, M.; Mertens, L. A.; Faure, A.; Carty, D.; Stoecklin, T.; Jankowski, P.; Szalewicz, K.; Le Picard, S. D.; Sims, I. R. Absolute Measurements of State-To-State Rotational Energy Transfer Between CO and $H_2$ at Interstellar Temperatures. *Physical Review A* **2022**, *105* (2), L020802. DOI: 10.1103/PhysRevA.105.L020802.

(69) Altkorn, R.; Zare, R. N. Effects of Saturation on Laser-Induced Fluorescence Measurements of Population and Polarization. *Annual Review of Physical Chemistry* **1984**, *35* (Volume 35), 265-289. DOI: https://doi.org/10.1146/annurev.pc.35.100184.001405.

(70) Western, C. M. PGOPHER: A Program for Simulating Rotational, Vibrational and Electronic Spectra. *Journal of Quantitative Spectroscopy and Radiative Transfer* **2017**, *186*, 221-242. DOI: https://doi.org/10.1016/j.jqsrt.2016.04.010.

(71) Masseron, T.; Plez, B.; Van Eck, S.; Colin, R.; Daoutidis, I.; Godefroid, M.; Coheur, P. F.; Bernath, P.; Jorissen, A.; Christlieb, N. CH in Stellar Atmospheres: An Extensive Linelist⋆. *Astronomy and Astrophysics* **2014**, *571*, 10.1051/0004-6361/201423956.

(72) Roberts, F. C.; Lehman, J. H. Infrared Frequency Comb Spectroscopy of $CH_2I_2$: Influence of Hot Bands and Pressure Broadening on the $v_1$ and $v_6$ Fundamental Transitions. *The Journal of Chemical Physics* **2022**, *156* (11), 114301. DOI: 10.1063/5.0081836 (acccessed 5/3/2024).

(73) Sadiek, I.; Hjältén, A.; Roberts, F. C.; Lehman, J. H.; Foltynowicz, A. Optical Frequency Comb-Based Measurements and the Revisited Assignment of High-Resolution Spectra of $CH_2Br_2$ in the 2960 to 3120 $cm^{-1}$ Region. *Physical Chemistry Chemical Physics* **2023**, *25* (12), 8743-8754, 10.1039/D2CP05881B. DOI: 10.1039/D2CP05881B.

(74) Gordon, I. E.; Rothman, L. S.; Hargreaves, R. J.; Hashemi, R.; Karlovets, E. V.; Skinner, F. M.; Conway, E. K.; Hill, C.; Kochanov, R. V.; Tan, Y.; Wcisło, P.; Finenko, A. A.; Nelson, K.; Bernath, P. F.; Birk, M.; Boudon, V.; Campargue, A.; Chance, K. V.; Coustenis, A.; Drouin, B. J.; Flaud, J. M.; Gamache, R. R.; Hodges, J. T.; Jacquemart, D.; Mlawer, E. J.; Nikitin, A. V.; Perevalov, V. I.; Rotger, M.; Tennyson, J.; Toon, G. C.; Tran, H.; Tyuterev, V. G.; Adkins, E. M.; Baker, A.; Barbe, A.; Canè, E.; Császár, A. G.; Dudaryonok, A.; Egorov, O.; Fleisher, A. J.; Fleurbaey, H.; Foltynowicz, A.; Furtenbacher, T.; Harrison, J. J.; Hartmann, J. M.; Horneman, V. M.; Huang, X.; Karman, T.; Karns, J.; Kassi, S.; Kleiner, I.; Kofman, V.; Kwabia–Tchana, F.; Lavrentieva, N. N.; Lee, T. J.; Long, D. A.; Lukashevskaya, A. A.; Lyulin, O. M.; Makhnev, V. Y.; Matt, W.; Massie, S. T.; Melosso, M.; Mikhailenko, S. N.; Mondelain, D.; Müller, H. S. P.; Naumenko, O. V.; Perrin, A.; Polyansky, O. L.; Raddaoui, E.; Raston, P. L.; Reed, Z. D.; Rey, M.; Richard, C.; Tóbiás, R.; Sadiek, I.; Schwenke, D. W.; Starikova, E.; Sung, K.; Tamassia, F.; Tashkun, S. A.; Vander Auwera, J.; Vasilenko, I. A.; Vigasin, A. A.; Villanueva, G. L.; Vispoel, B.; Wagner, G.; Yachmenev, A.; Yurchenko, S. N. The HITRAN2020 Molecular Spectroscopic Database. *Journal of Quantitative Spectroscopy and Radiative Transfer* **2022**, *277*, 107949. DOI: https://doi.org/10.1016/j.jqsrt.2021.107949.

(75) Fayt, A.; Vandenhaute, R.; Lahaye, J. G. Global Rovibrational Analysis of Carbonyl Sulfide. *Journal of Molecular Spectroscopy* **1986**, *119* (2), 233-266. DOI: https://doi.org/10.1016/0022-2852(86)90022-6.

(76) Lahaye, J. G.; Vandenhaute, R.; Fayt, A. $CO_2$ Laser Saturation Stark Spectra and Global Rovibrational Analysis of the Main Isotopic Species of Carbonyl Sulfide ($OC^{34}S$, $O^{13}CS$, and $^{18}OCS$). *Journal of Molecular Spectroscopy* **1987**, *123* (1), 48-83. DOI: https://doi.org/10.1016/0022-2852(87)90262-1.

(77) Régalia-Jarlot, L.; Hamdouni, A.; Thomas, X.; Von der Heyden, P.; Barbe, A. Line Intensities of the: $v_3$, $4v_2$, $v_1+v_3$, $3v_1$ and $2v_1+2v_2$ Bands of $^{16}O^{12}C^{32}S$ Molecule. *Journal of Quantitative Spectroscopy and Radiative Transfer* **2002**, *74* (4), 455-470. DOI: https://doi.org/10.1016/S0022-4073(01)00267-9.

(78) Bogey, M.; Bauer, A. Microwave Spectroscopy of OCS in Highly Excited Vibrational States Through Energy Transfer from $N_2$. *Journal of Molecular Spectroscopy* **1980**, *84* (1), 170-178. DOI: https://doi.org/10.1016/0022-2852(80)90251-9.

(79) Yamada, K. M. T.; Klebsch, W. High-Temperature Spectrum of OCS in a dc Discharge by Diode Laser Spectroscopy. *Journal of Molecular Spectroscopy* **1987**, *125* (2), 380-392. DOI: https://doi.org/10.1016/0022-2852(87)90105-6.

(80) Koshelev, M. A.; Tretyakov, M. Y. Collisional Broadening and Shifting of OCS Rotational Spectrum Lines. *Journal of Quantitative Spectroscopy and Radiative Transfer* **2009**, *110* (1), 118-128. DOI: https://doi.org/10.1016/j.jqsrt.2008.09.010.

(81) Harris, S. J.; Janda, K. C.; Novick, S. E.; Klemperer, W. Intermolecular Potential between an Atom and a Linear Molecule: The Structure of ArOCS *The Journal of Chemical Physics* **1975**, *63* (2), 881-884. DOI: 10.1063/1.431368 (acccessed 5/3/2024).

(82) Lovas, F. J.; Suenram, R. D. Pulsed Beam Fourier Transform Microwave Measurements on OCS and Rare Gas Complexes of OCS with Ne, Ar, and Kr. *The Journal of Chemical Physics* **1987**, *87* (4), 2010-2020. DOI: 10.1063/1.453176 (acccessed 5/3/2024).

(83) Hayman, G. D.; Hodge, J.; Howard, B. J.; Muenter, J. S.; Dyke, T. R. Infrared Absorption Spectra of Ar-OCS and Kr-OCS van der Waals Complexes in the Carbonyl Stretching Region. *Journal of Molecular Spectroscopy* **1989**, *133* (2), 423-437. DOI: https://doi.org/10.1016/0022-2852(89)90202-6.

(84) Hayman, G. D.; Hodge, J.; Howard, B. J.; Muenter, J. S.; Dyke, T. R. Molecular-Beam Infrared Absorption Studies of Rare Gas-OCS Complexes. *Chemical Physics Letters* **1985**, *118* (1), 12-18. DOI: https://doi.org/10.1016/0009-2614(85)85256-8.

(85) Zhu, H.; Guo, Y.; Xue, Y.; Xie, D. Ab Initio Potential Energy Surface and Predicted Microwave Spectra for Ar-OCS Dimer and Structures of $Ar_n$-OCS (n = 2–14) Clusters. *Journal of Computational Chemistry* **2006**, *27* (9), 1045-1053. DOI: https://doi.org/10.1002/jcc.20421 (acccessed 2024/05/03).

(86) Sun, C.; Wang, Z.; Feng, E.; Zhang, C. A Three-Dimensional Potential Energy Surface and Infrared Spectra for the Ar–OCS van der Waals Complex. *Chemical Physics Letters* **2014**, *592*, 182-187. DOI: https://doi.org/10.1016/j.cplett.2013.12.053.

(87) Bone, R. G. A. An Investigation of the Structure of Weakly Bound $(OCS)_2$. *Chemical Physics Letters* **1993**, *206* (1), 260-270. DOI: https://doi.org/10.1016/0009-2614(93)85550-8.

(88) Afshari, M.; Dehghani, M.; Abusara, Z.; Moazzen-Ahmadi, N.; McKellar, A. R. W. Observation of the "Missing" Polar OCS Dimer. *The Journal of Chemical Physics* **2007**, *126* (7), 071102. DOI: 10.1063/1.2709879 (acccessed 5/3/2024).

(89) Minei, A. J.; Novick, S. E. Microwave Observation of the "Recently Found" Polar OCS Dimer. *The Journal of Chemical Physics* **2007**, *126* (10), 101101. DOI: 10.1063/1.2715544 (acccessed 5/3/2024).

(90) Miller, I.; Faulkner, T.; Saunier, J.; Raston, P. L. Observation of the Elusive "Oxygen-in" OCS Dimer. *The Journal of Chemical Physics* **2020**, *152* (22), 221102. DOI: 10.1063/5.0010716 (acccessed 5/3/2024).

(91) Randall, R. W.; Wilkie, J. M.; Howard, B. J.; Muenter, J. S. Infrared Vibration-Rotation Spectrum and Structure of OCS Dimer. *Molecular Physics* **1990**, *69* (5), 839-852. DOI: 10.1080/00268979000100641.

(92) Zabarnick, S.; Fleming, J. W.; Lin, M. C. Kinetics of CH Radical Reactions with $N_2O$, $SO_2$, OCS, $CS_2$, and $SF_6$. *International Journal of Chemical Kinetics* **1989**, *21* (9), 765-774. DOI: 10.1002/kin.550210905.

(93) Lucas, D. I.; Kavaliauskas, C. J.; Blitz, M. A.; Heard, D. E.; Lehman, J. H. Ab Initio and Statistical Rate Theory Exploration of the CH (X $^2\Pi$) + OCS Gas-Phase Reaction. *The Journal of Physical Chemistry A* **2023**, *127* (31), 6509-6520. DOI: 10.1021/acs.jpca.3c01082.

(94) Le Picard, S. D.; Canosa, A. Measurement of the Rate Constant for the Association Reaction CH + $N_2$ at 53 K and its Relevance to Triton's Atmosphere. *Geophysical Research Letters* **1998**, *25* (4), 485-488. DOI: https://doi.org/10.1029/98GL50118 (acccessed 2024/05/03).

(95) D. Le Picard, S. b.; Canosa, A.; R. Rowe, B.; A. Brownsword, R.; W. M. Smith, I. Determination of the Limiting Low Pressure Rate Constants of the Reactions of CH with $N_2$ and CO: a CRESU Measurement at 53 K. *Journal of the Chemical Society, Faraday Transactions* **1998**, *94* (19), 2889-2893, 10.1039/A803930E. DOI: 10.1039/A803930E.